\documentclass[10pt,journal]{IEEEtran}

\usepackage{amsfonts}
\usepackage{amsmath}
\usepackage{amssymb}
\usepackage{amsthm}
\usepackage{graphicx}
\usepackage{subfigure}
\usepackage{mathdefs}
\usepackage{mathrsfs}
\usepackage{lineno}
\usepackage{enumitem}

\usepackage[noend]{algpseudocode}
\usepackage{algorithm}
\usepackage{placeins}
\usepackage{bm}
\usepackage{kbordermatrix}
\usepackage{todonotes}
\usepackage{enumitem}
\usepackage{booktabs}
\usepackage{verbatimbox}

\usepackage{dsfont}
\usepackage{colortbl}
\usepackage{mathtools}

\usepackage{upgreek}

\definecolor{lg}{gray}{0.9}

\newtheorem{corollary}{Corollary}
\newtheorem{definition}{Definition}

\newtheorem{fact}{Fact}
\newtheorem{lemma}{Lemma}
\newtheorem{proposition}{Proposition}
\newtheorem{theorem}{Theorem}

\newcommand{\corref}[1]{Cor.~\ref{corr:#1}}
\newcommand{\defref}[1]{Def.~\ref{def:#1}}

\newcommand{\facref}[1]{Fact~\ref{fac:#1}}
\newcommand{\lemref}[1]{Lem.~\ref{lem:#1}}
\newcommand{\prpref}[1]{Prop.~\ref{prp:#1}}
\newcommand{\thmref}[1]{Thm.~\ref{thm:#1}}
\newcommand{\eqnref}[1]{(\ref{eq:#1})}
\newcommand{\secref}[1]{\S\ref{sec:#1}}
\newcommand{\appref}[1]{App.~\ref{app:#1}}
\newcommand{\figref}[1]{Fig.~\ref{fig:#1}}
\newcommand{\tabref}[1]{Table~\ref{tab:#1}}

\usepackage{hyperref}
\hypersetup{colorlinks=false}
\hypersetup{colorlinks,citecolor=black,filecolor=black,linkcolor=black,urlcolor=black}

\begin{document}

\title{Graph search via star sampling with and without replacement}
\author{Jonathan Stokes and Steven Weber
\thanks{Preliminary published versions of this work include \cite{StoWeb2016b,StoWeb2017a,StoWeb2017b}.  This work was supported by the National Science Foundation $\#$IIS-1250786. J.\ Stokes was and S.\ Weber is with the Department of Electrical and Computer Engineering of Drexel University.  S.\ Weber is the contact author: {\sf sweber@coe.drexel.edu}.}}

\maketitle

\begin{abstract}
Star sampling (SS) is a random sampling procedure on a graph wherein each sample consists of a randomly selected vertex (the star center) and its (one-hop) neighbors (the star points).  We consider the use of SS to find any member of a target set of vertices in a graph, where the figure of merit (cost) is either the expected number of samples (unit cost) or the expected number of star centers plus star points (linear cost) until a vertex in the target set is encountered, either as a star center or as a star point.  We analyze these two performance measures on three related star sampling paradigms: SS with replacement (SSR), SS without center replacement (SSC), and SS without star replacement (SSS).  Exact and approximate expressions are derived for the expected unit and linear costs of SSR, SSC, and SSS on Erd\H{o}s-R\'{e}nyi (ER) random graphs. The approximations are seen to be accurate.  SSC/SSS are notably better than SSR under unit cost for low-density ER graphs, while SSS is notably better than SSR/SSC under linear cost for low- to moderate-density ER graphs.  Simulations on twelve ``real-world'' graphs shows the cost approximations to be of variable quality: the SSR and SSC approximations are uniformly accurate, while the SSS approximation, derived for an ER graph, is of variable accuracy.
\end{abstract}


\section{Introduction}
\label{sec:intro}
Let $G = (V,E)$ denote a simple undirected graph with vertex set $V$ and edge set $E$.  Suppose each vertex has a property value, defined via a function $f : V \to P$, for $P$ the set of property values.  Then $V^* \equiv f^{-1}(P^*)$, for $P^* \subset P$, is the subset of vertices holding property values of interest (i.e., vertices with properties in $P^*)$.  This paper evaluates the performance of three related random sampling approaches to finding a vertex in $V^*$, called {\em star sampling}, described below, that differ in terms of which part of the sample is replaced.  

{\em Star sampling.} Using random sampling to search for a vertex of interest is often suitable for large and/or dynamic graphs, where either the order/size and/or the rapid evolution of the graph precludes holding the graph in local memory.  In such cases the searcher may be required to {\em query} the graph, by requesting a random vertex.  {\em Star sampling} (SS) is a variant on vertex sampling in which each sample returns not only the property value $f(v)$ of the selected vertex $v \in V$, termed the {\em star center}, but also the property values $f(N_G(v))$ of its (one-hop) neighbors $N_G(v)$ (where $N_G(v) \equiv \{u \in V : uv \in E\}$ is the neighborhood of $v$, and $uv \in E$ denotes vertices $u$ and $v$ are joined by an edge), termed the {\em star points}.  Random star sampling selects a vertex uniformly at random, then checks whether either it or any of its neighbors hold the property of interest; this is repeated until a vertex in $V^*$ is found.

{\em Cost.} Although large graphs are encoded in a variety of ways, it is often the case that the data structure corresponding to each vertex (e.g., the profile of a particular member in a social network) includes the list of neighbors of that vertex (e.g., the social connections of that member in the network).  Star sampling is a practical sampling paradigm whenever such neighbor information is available.  
The property may be readily available if it is easily computable, or if it is not easily computable, but has been {\em precomputed} and stored in the data structure for the vertex.  To address this issue we consider two natural cost models ({\em unit} and {\em linear}), where cost is measured either as the number of star samples (unit cost) or as the number of vertices (linear cost) for which property values are queried / computed.  Unit cost is most natural for the case when the primary cost incurred is the query of the star itself and property values of neighbors are readily available, while linear cost is most natural for the case where the primary cost incurred is the computation of the property value.  

{\em Target set.} We focus on the case where $P^*$ is independent of $G$, i.e., the target set is a subset $\Vsf^*$, chosen uniformly at random from $V$, with prescribed cardinality $|\Vsf^*| = n^*$.  As such, the expected search cost to find a member of the target set depends upon the set only through $n^*$.

{\em Variants.} Three SS variants are considered:
\begin{itemize}[leftmargin=*]
\item {\em SS with replacement (SSR)}: the star center is selected uniformly at random from the set of vertices;
\item {\em SS without center replacement (SSC)}: the star center is selected uniformly at random from the set of {\em remaining} vertices; the star center (along with its adjacent edges) is removed from the graph after the query;
\item {\em SS without star replacement (SSS)}: the star center is selected uniformly at random from the set of {\em remaining} vertices; the entire star (center, points, and all adjacent edges) is removed from the graph after the query.
\end{itemize}

{\em Urn sampling.} One motivation in considering these variants is to understand their relative performance, in a manner similar to the elementary case of sampling balls from an urn.  When seeking any one of $n^*$ marked balls out of a total of $n \geq n^*$ balls in an urn, sampling {\em with} replacement requires on average $n/n^*$ samples; this follows immediately from the observation that the number of draws until the first success, say $\csf^{\rm R}$, is a geometric random variable (RV) with success probability $n^*/n$, and expectation $\Ebb[\csf^{\rm R}] = n/n^*$.  In contrast, sampling {\em without} replacement requires a random number of draws, $\csf^{\rm NR}$, with expectation given by the following Fact.

\begin{fact}
\label{fac:ecnra}
$\Ebb[\csf^{\rm NR}] = (n+1)/(n^*+1)$. 
\end{fact}

The proof is found in \appref{ecnra}.

Thus, the performance ratio of the expected number of samples with vs.\ without replacement is $\Ebb[\csf^{\rm R}]/\Ebb[\csf^{\rm NR}] = (1+1/n^*)/(1+1/n)$.  For $n^* \ll n$, sampling without replacement improves the mean search time by at most two, relative to sampling with replacement, with equality for $n^*=1$.


{\em Contributions.} The three primary contributions are:
\begin{itemize}[leftmargin=*]
\item Derivation of exact, bound, and approximate expected unit and linear cost expressions, for both arbitrary graphs and Erd\H{o}s-R\'{e}nyi (ER) random graphs, for the three star sampling variants; these results are summarized in \tabref{resultssummary}.
\item Numerical assessment of quality, in which the bounds are shown to be tight, and the approximations are seen to be accurate, relative to simulation results of ER random graphs; these results are seen in \secref{ucnr} (\figref{ucnr}) and \secref{lcnr} (\figref{lcnr}).
\item Numerical assessment on ``real-world'' graphs, in which the relative error of the approximations are compared with simulation results; c.f.\ \secref{results} (\tabref{unit_MD_RND} and \tabref{lin_MD_RND}). 
\end{itemize}
Three points of context for the results bear mention.  
\begin{itemize}[leftmargin=*]
\item The accuracy of the cost approximations is notable, particularly for SSS, given the complexity of the dynamics under star sampling without star replacement.  The conditional probability of hitting the star at time $t$, conditioned on not hitting it before, is approximated by tracking both the expected reduction in the overall order of the graph and the expected reduction in the number of {\em neighbors} of the target set in the previous $t-1$ samples.  This yields an approximate {\em unconditional} probability of first hitting the target set at $t$, which in turn yields an approximate expected cost.  
\item The (mostly) small relative error of the cost approximations for ``real-world'' graphs is notable, particularly for SSS, given the approximations are derived for ER graphs, which are quite distinct from the ``real-world'' graphs.  
\item It is notable that all cost approximations require knowing only {\em three} quantities: the graph order ($n$), the target set cardinality ($n_0^*$), and the graph edge density $s$.
\end{itemize}

\begin{table}
\begin{center}
\begin{tabular}{llll}
Graph & Variant & Unit cost (\secref{unit_cost}) & Linear cost (\secref{s_cost}) \\ \hline \hline
Arbitrary & SSR & \facref{ssr} (exact) & \facref{lcagssr} (exact) \\
Arbitrary & SSC & \facref{ssc} (exact) & N/A \\
Arbitrary & SSS & N/A & N/A \\ \hline
Erd\H{o}s-R\'{e}nyi & SSR & \prpref{unicostssrsscer} (bounds) & \prpref{lcssrer} (approx.) \\
Erd\H{o}s-R\'{e}nyi & SSC & \prpref{unicostssrsscer} (bounds) & \prpref{lcsscer} (approx.) \\
Erd\H{o}s-R\'{e}nyi & SSS & \thmref{appuncprobhittarsetSSS} (approx.) & \prpref{lcssser} (approx.) \\ \hline
\end{tabular}
\label{tab:resultssummary}
\caption{Summary of results}
\end{center}
\end{table}

{\em Outline.} The paper is organized as follows. \secref{related} discusses related work. \secref{model} provides basic notation and definitions. \secref{unit_cost} and \secref{s_cost} study performance under the {\em unit} and {\em linear} cost models, respectively. \secref{results} assesses the accuracy and performance of the cost estimates on ``real-world'' graphs. \secref{conclusion} holds discussion and conclusion.  Most proofs are in the Appendices.

\section{Related work}
\label{sec:related}

There is an extensive literature on graph sampling, too vast to credibly review here. Classic graph exploration strategies include: {\em random sampling} of vertices or edges, {\em random walks}, and {\em random jump sampling}, which alternates between a {\em random walk} and {\em random sampling}. These graph exploration strategies are general in the sense that they may be applied towards a variety of objectives.  Two notable classes of objectives include $i)$ sampling to derive an unbiased estimate of some graph property, and $ii)$ sampling to search for vertices with particular properties; this paper falls into the latter category.  


{\em Snowball sampling} \cite{Kol2009} was introduced in \cite{Goo1961} (see also \cite{Fra1977,LeeKim2006,AhnHan2007,HuLau2013}); {\em star sampling}, a special case of snowball sampling, appears in \cite{WanLu2013}.  Sampling bias is addressed in \cite{StuRej2009,MaiWol2011} and sampling bias for star sampling is addressed in \cite{DasKum2012}.  Ref.\ \cite{LeeXu2012} argues that Metropolis-Hastings sampling algorithms should avoid backtracking. Ref.\ \cite{LiYu2015} proposed a \textit{rejection controlled Metropolis-Hastings} algorithm and a \textit{non-backtracking generalized maximum-degree sampling} algorithm.  Other graph sampling algorithms include {\em albatross sampling} \cite{JinChe2011} and {\em rank degree} \cite{VouSal2016}.  Ref.\ \cite{GjoKur2010} found that a \textit{Metropolis-Hasting random walk} and a \textit{re-weighted random walk} both outperform a simple random walk in returning a uniform sample of Facebook users.  Ref.\ \cite{KurGjo2011} has shown that \textit{weighted random walks} can be used to carry out stratified sampling on graphs, while \cite{AvrBor2018} develops random walk sampling techniques that avoid the ``burn-in'' period to reach stationarity (c.f.\ \cite{LvCao2002,IkeKub2003,GkaMih2006,CooFri2007,AviKri2008,AvrRib2010,RibTow2010} for additional earlier relevant work on using random walks for graph search). Ref.\ \cite{ChiDas2016} bounds the number of steps required to return a uniform sample of a network using \textit{rejection sampling}, \textit{maximum-degree sampling}, and \textit{Metropolis-Hastings sampling}.  Graph search for power-law (``scale--free'') networks is addressed in \cite{AdaLuk2001}.  Graph search for large degree vertices is the focus of \cite{BraKea2010,AvrLit2012,CooRad2012,AvrLit2014a,AvrLit2014b,CooRad2014}.  Finally, sampling with vs.\ without replacement is addressed in \cite{WhiAnd1982}.
 
Our pertinent prior work includes \cite{StoWeb2016b,StoWeb2017a,StoWeb2017b}. Ref.\ \cite{StoWeb2016b} compares the cost of using random walk vs.\ random sampling to find a target vertex.  Ref.\ \cite{StoWeb2017a} analyzes the cost of finding a degree $j$ vertex or an edge with vertices of degrees $j,k$ via star sampling without replacement.  Ref. \cite{StoWeb2017b} holds preliminary results comparing star sampling with and without replacement. To our knowledge, ours is the first work to study the performance impact of replacement on star sampling.  Additional related prior work includes \cite{StoWeb2016a,StoWeb2016c,StoWeb2018}.

\section{Notation, Sampling Model, Background}
\label{sec:model}

Notation is defined in \secref{notation}, ER graphs and their properties are described in \secref{ergp}, and the three sampling models and two cost models are stated in \secref{samplingmodel}.

\subsection{Notation}
\label{sec:notation}

Let $a \equiv b$ denote $a$ and $b$ are equal by definition.  Denote the set $\{1,2,3,\ldots\}$ by $\Nbb$ and $\{0,1,2,\ldots,\}$ by $\Zbb_+$.  Let $[n]$ denote $\{1,\ldots,n\}$, for $n \in \Nbb$.  Scalar random variables are denoted in a lowercase sans-serif font, e.g., $\xsf,\ksf$, while graph- and set-valued random variables are denoted with uppercase sans-serif font, e.g., $\Gsf,\Vsf,\Esf$. Expectation is denoted $\Ebb[\cdot]$, probability is denoted $\Pbb(\cdot)$, and IID refers to independent and identically distributed.  Four probability distributions are used: $i)$ $\mathrm{uni}(U)$ denotes a RV taking value uniformly over set $U$, $ii)$ $\mathrm{Ber}(p)$ denotes a Bernoulli RV with $p \in [0,1]$ and support $\{0,1\}$, $iii)$ $\mathrm{bin}(n,p)$ denotes a binomial RV with parameters $(n,p)$, for $n \in \Nbb$ and $p \in [0,1]$, and support $\{0,1,2,\ldots,n\}$, and $iv)$ $\mathrm{geo}(p)$ denotes a geometric RV with $p \in [0,1]$ and support $\Nbb$. 

The following graph and sampling notation is used:
\begin{itemize}[leftmargin=*]

\item {\em Order, size, edges.} An undirected and simple graph of order $n$ is denoted $G = (V,E)$, with vertex set $V \equiv [n]$ and edge set $E$; size is denoted by $m \equiv |E|$. An edge is denoted $uv$.  

\item {\em Complete, empty, order-$0$ graphs.}  $K_n$ denotes the {\em complete} graph of order $n$, with all $m = \binom{n}{2}$ edges present.  Let $\bar{G} = (V,\bar{E})$ denote the {\em complement} of $G$, i.e., the edge set $\bar{E}$ of $\bar{G}$ is the complement of the edge set $E$ of $G$.  Then $\bar{K}_n$ denotes the {\em empty} graph of order $n$, with $m = 0$ edges present.  The {\em order-$0$ graph}, $\bar{K}_0$, has $n=0$ vertices, i.e., $V = E = \emptyset$.  

\item {\em Neighborhoods.} Let $N_G(v) \equiv \{u \in V : uv \in E\}$ denote the (direct) neighbors of $v$, $N_G^e(v) \equiv N_G(v) \cup \{v\}$ the {\em extended} neighborhood of $v$, $\Gamma_G(v) \equiv \{ uv \in E\}$ the edge neighborhood of $v$, i.e., the edges adjacent to $v$, and 
\begin{equation}
\Gamma^e_G(v) \equiv \bigcup_{u \in N_G(v)} \Gamma_G(u),
\end{equation}
the {\em extended} edge neighborhood of $v$, i.e., all edges adjacent to $v$ or any of $v$'s neighbors.  Observe $N_G^e(v)$ is a star sample with star center $v$ and star points $N_G(v)$.  For $V^* \subseteq V$, let $N_G(V^*) \equiv (\bigcup_{v \in V^*} N_G(v)) \setminus V^*$ denote neighbors of $V^*$ not including $V^*$, and let $N_G^e(V^*) \equiv \bigcup_{v \in V^*} N_G^e(v)$ denote $V^*$ and its neighbors.  

\item {\em Degrees.} Let $d_G(v) \equiv |N_G(v)|$ denote the degree of $v$, and $d_G^e(v) \equiv |N_G^e(v)|$ the ``extended degree'', i.e., $d_G^e(v) = d_G(v) + 1$.  Let $D_G \equiv \bigcup_{v \in V} d_G(v)$ denote the set of degrees found in $G$.  Partition $V$ by degree into subsets $(V_G(k), k \in D)$, with $V_G(k) \equiv \{ v \in V : d_G(v) = k\}$ the set of vertices with degree $k$ and $w_G(k) \equiv |V_G(k)|/n$ the fraction of vertices with degree $k$, to obtain the degree {\em distribution} of $G$, denoted $(w_G(k), k\in D)$ (with $\sum_{k \in D} w_G(k) = 1$).   The {\em expected degree} 
of a randomly selected $\vsf \sim \mathrm{uni}(V)$ is 
\begin{equation}
d_G \equiv \Ebb[d_G({\vsf})] = \sum_{k \in D_G} k w_G(k) = \frac{1}{n} \sum_{v \in V} d_G(v).
\end{equation}
\end{itemize}
The notation in this subsection is summarized in \tabref{notation}.

\begin{table}
\centering
\begin{tabular}{ll}
Notation & Description \\ \hline \hline
$d_G(v)$ & degree of vertex $v$ \\
$d_G^e(v)$ & extended degree of vertex $v$ \\
$d_G$ & average degree of $G$, i.e., $\Ebb[d_G(\vsf)]$, for $\vsf \sim \mathrm{uni}(V)$ \\
$E$ & edge set \\
$G$ & simple undirected graph \\
$\Gamma_G(v)$ & edge neighborhood of $v$ \\
$\Gamma^e_G(v)$ & extended edge neighborhood of $v$ \\
$m$ & graph size, $m=|E|$ \\
$N_G(v)$ & neighbors of vertex $v$ \\
$N_G(V^*)$ & neighbors of target set $V^*$ \\
$N^e_G(v)$ & extended neighbors of vertex $v$ \\
$N^e_G(V^*)$ & extended neighbors of target set $V^*$ \\
$n$ & graph order, $n=|V|$ \\
$n_G^{e,*}$ & order of the extended target set, $n_G^{e,*} = |N_G^e(V^*)|$ \\
$s$ & edge density, i.e., $s = |E|/\binom{n}{2}$ \\
$uv$ & edge between vertices $u$ and $v$ is in $E$ \\
$V$ & vertex set \\
$V^*$ & target set \\
$V_G(k)$ & set of vertices in $G$ of degree $k$ \\
$w_G(k)$ & fraction of vertices in $G$ of degree $k$ \\
\hline
\end{tabular}
\label{tab:notation}
\caption{Select notation from \secref{notation}.}
\end{table}

\subsection{Erd\H{o}s R\'enyi (ER) graph properties}
\label{sec:ergp}

An Erd\H{o}s-R\'{e}nyi (ER) random graph $\Gsf = (V,\Esf)$ has parameters $(n,s)$, where $n \in \Nbb$ denotes the order, $V = [n]$, and $s \in (0,1)$ denotes the edge probability.  A realization of an ER random graph, i.e., of the random edge set $\Esf$, is obtained by including each of the $\binom{n}{2}$ possible edges independently at random with probability $s$.  The random size of $\Gsf$, denoted $\msf \equiv |\Esf|$, is a binomial RV, i.e., $\msf \sim \mathrm{bin}(\binom{n}{2},s)$, as is the degree $d_{\Gsf}(\vsf)$ of a randomly selected vertex $\vsf\sim \mathrm{uni}(V)$, namely, $d_{\Gsf}(\vsf) \sim \mathrm{bin}(n-1,s)$.  The RV $\ssf \equiv \msf/\binom{n}{2}$ is the {\em edge density} (with $\Ebb[\ssf] = s$), and the RV $\dsf \equiv (n-1) \ssf = 2 \msf / n$ is the {\em average degree} of $\Gsf$.  Given an ER graph $\Gsf$, the RV $\dsf$ is the expected degree of a vertex selected uniformly at random.  Although unconventional, it is useful to extend the domain of $n$ from $\Nbb$ to $\Zbb_+$, i.e., allowing $n = 0$, so that the order-$0$ graph $\bar{K}_0$ (with $V = E = \emptyset$) is a {\em trivial} ER graph.

\subsection{Sampling variant definitions and cost models}
\label{sec:samplingmodel}

Random star sampling from a given graph $G = (V,E)$ produces a sequence of random graphs, denoted $(\Gsf_t, t \in \Nbb)$, where $\Gsf_t = (\Vsf_t,\Esf_t)$ is the random graph after sample $t$.  Both the vertex set and edge set are (in general) random variables.  It is convenient to denote the given graph as $G_0 = (V_0,E_0)$, i.e., the initial member of the list of random graphs, corresponding to $t=0$.  The target set is denoted by $V^*$ or $V_0^*$.  Define the sequence of random vertex sets $(\Vsf^*_t, t \in \Nbb)$, where $\Vsf^*_t \equiv \Vsf_t \cap V_0^*$ holds the members of the initial target set still ``alive'' after $t$ samples.  The sampling construction ensures the random sets are nested: $\Vsf_{t+1} \subseteq \Vsf_t$, $\Esf_{t+1} \subseteq \Esf_t$, and $\Vsf^*_{t+1} \subseteq \Vsf^*_t$.  Recalling \secref{intro}, the three SS variants are as follows.  

\begin{definition}
SS with replacement (SSR).  Generate the IID random sequence of star centers $(\vsf_t, t \in \Nbb)$, with $\vsf_t \sim \mathrm{uni}(V_0)$.  As SSR uses replacement, $\Gsf_t = G_0$ for all $t$.  
\end{definition}

\begin{definition}
SS without center replacement (SSC).  Generate the random sequence of star centers $(\vsf_t, t \in [n])$, with $\vsf_t \sim \mathrm{uni}(\Vsf_{t-1})$, and update the graph by removing the star center, i.e., $\Vsf_t = \Vsf_{t-1} \setminus \{\vsf_t\}$, and the edges in the edge neighborhood of the star center, i.e., $\Esf_t = \Esf_{t-1} \setminus \Gamma_{\Gsf_{t-1}}(\vsf_t)$.  
\end{definition}

\begin{definition}
SS without star replacement (SSS).  Generate the random sequence of star centers $(\vsf_t, t \in [n])$, with $\vsf_t \sim \mathrm{uni}(\Vsf_{t-1})$, and update the graph by removing the star, i.e., $\Vsf_t = \Vsf_{t-1} \setminus N^e_{\Gsf_{t-1}}(\vsf_t)$, and the edges in the {\em extended} edge neighborhood of the star center, i.e., $\Esf_t = \Esf_{t-1} \setminus \Gamma^e_{\Gsf_{t-1}}(\vsf_t)$. 
\end{definition}

We focus our analysis of star sampling on ER graphs since they are closed under both SSC and SSS, i.e., if either a star-center (SSC) or a star (SSS) is removed from an ER graph the resulting graph is still an ER graph, albeit one with a different (reduced) order.  This follows intuitively from the fact that the presence or absence of an edge is independent across edges, but is established formally in \lemref{tbd1} below.  It is also possible that the graph resulting from an SSC or SSC operation is {\em trivial}, i.e., the order-$0$ graph $\bar{K}_0$ (c.f., \secref{notation} and \secref{ergp}).

\begin{lemma}
\label{lem:tbd1}
Given an ER random graph, the graph that remains after an SSS or SSC random sample is an ER random graph.
\end{lemma}

The proof is found in \appref{tbd1}.

\lemref{gosm} gives the expected number of edges removed from an ER graph under SSS. Consider a star sample with star center $\vsf \sim \mathrm{uni}(V)$ of an ER graph $\Gsf$ with parameters $(n,s)$; recall $\vsf$ has degree distribution $\dsf \sim \mathrm{bin}(n-1,s)$.

\begin{lemma}
\label{lem:gosm}
Given an ER random graph $\Gsf$ with parameters $(n,s)$, a randomly selected vertex $\vsf$ as star center, and conditioned on the degree $\dsf$ of $\vsf$, the random number of edges in the extended edge neighborhood of $\vsf$, denoted $\gsf \equiv |\Gamma^e_{\Gsf}(\vsf)|$, has a shifted binomial distribution
\begin{equation}
\label{eq:gsfgdsfnumedgesss}
\gsf|\dsf \sim \dsf + \mathrm{bin}\left(\binom{\dsf}{2} + \dsf(n-\dsf-1),s\right),
\end{equation}
with (unconditional) expectation 
\begin{equation}
\Ebb[\gsf] = (n-1)s(1+(n/2-1)(2-s)s). \label{eq:expgsfnumedgesss}
\end{equation}
The asymptotic (in $n$) ratio of $\Ebb[\gsf]$ to the expected total number of edges in the graph $\binom{n}{2} s$ in the graph is $\lim_{n \to \infty} \Ebb[\gsf]/(\binom{n}{2}s) = (2-s)s$.
\end{lemma}

\begin{figure}[!ht]
\centering
\includegraphics[width=\columnwidth]{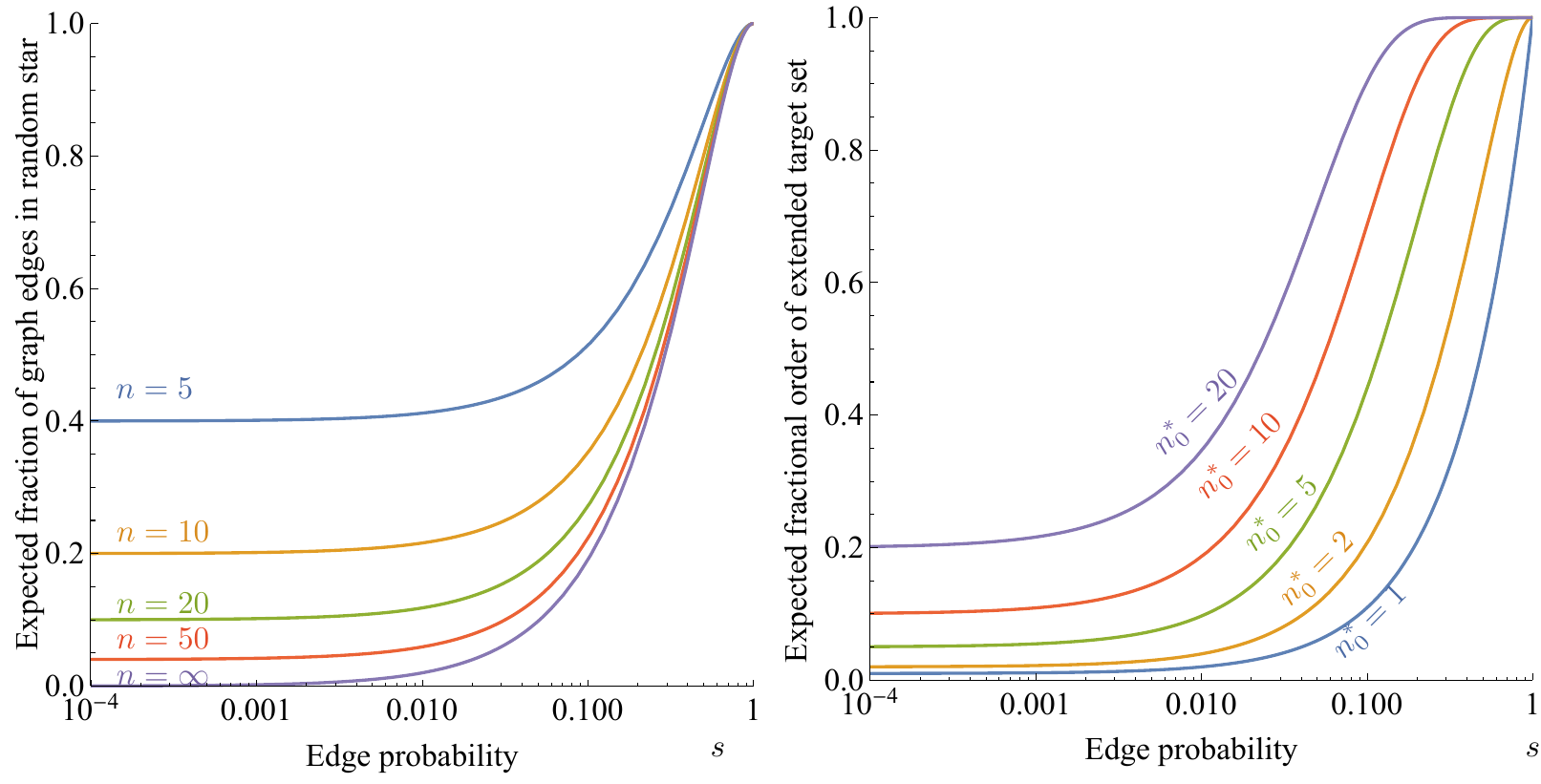}
\caption{{\em Left:} expected fraction of ER graph edges in a star sample vs.\ edge probability $s$, for $n \in \{5,10,20,50\}$ and $\lim n \uparrow \infty$, i.e., $(2-s)s$ (\lemref{gosm}).
{\em Right:} Expected fraction of vertices from an ER random graph with parameters $(n=100,s)$ in the extended neighborhood, i.e., $n_0^{e,*}/n$ vs.\ the edge probability $s$, for $n_0^* \in \{1,2,5,10,20\}$ (\facref{nsf0esER}).}
\label{fig:gosm}
\end{figure}

The proof is found in \appref{gosm}.  See \figref{gosm} (left).

The unit and linear costs for SS are defined below.

\begin{definition}
\label{def:perf} 
The {\em unit cost} of a SS is the random number of samples until a star, either the star center or one of the star points, intersects the target set $V^*$, i.e., 
\begin{equation}
\label{eq:perfdef}
\csf_u(G,V^*) \equiv \min\{ t \in \Nbb : N^e_{\Gsf_{t-1}}(\vsf_t) \cap V^* \neq \emptyset\}.
\end{equation}
This cost is a function of the sequence of random graphs $(\Gsf_t, t \in \Nbb)$.  The {\em expected unit cost} of a SS, denoted 
\begin{equation}
\label{eq:perfdef2}
c_u(G,V^*) \equiv \Ebb[\csf_u(G,V^*)],
\end{equation}
is the expectation of $\csf_u(G,V^*)$, taken with respect to the distribution over all possible realizations of graph sequences induced by SS that begin with $G_0 = G$.
\end{definition}

\begin{definition}
\label{def:perf2} 
The {\em linear cost} of a SS is the random sum of the extended degrees of the randomly selected vertex centers from each star sample until a star, either the star center or one of the star points, intersects the target set $V^*$, i.e., 
\begin{equation}
\label{eq:perfdef3}
\csf_l(G,V^*) \equiv \sum_{t=1}^{\csf_u(G,V^*)} d_{\Gsf_{t-1}}^e(\vsf_t).
\end{equation}
This cost is a function of the sequence of random graphs $(\Gsf_t, t \in \Nbb)$.  The {\em expected linear cost} of a SS, denoted 
\begin{equation}
\label{eq:perfdef4}
c_l(G,V^*) \equiv \Ebb[\csf_l(G,V^*)],
\end{equation}
is the expectation of $\csf_l(G,V^*)$, taken with respect to the distribution over all possible realizations of graph sequences induced by SS that begin with $G_0 = G$.
\end{definition}

\section{Unit cost model}
\label{sec:unit_cost}

Expected unit costs of SSR and SSC are given in \secref{ucrc} and the expected unit cost of SSS is given in \secref{ucs}.  Exact results in \secref{ucrc} are given for an arbitrary graph and as bounds for ER random graphs, while approximate results in \secref{ucs} are only given for ER random graphs.  \secref{ucnr} gives numerical results.

\subsection{Unit cost analysis for SSR and SSC}
\label{sec:ucrc}

Let $G = (V,E)$ be an arbitrary graph of order $n$, and let $V^* \subseteq V$ be an arbitrary target set. Recall from \secref{notation} that $N^e_G(V^*)$ contains the target set $V^*$ and its neighbors in $G$.  Observe the equivalence: a star sample $N^e_G(v)$ intersects $V^*$ if and only if $v \in N^e_G(V^*)$. This observation yields the expected unit cost of SSR and SSC.  Set $n^{e,*}_G \equiv |N^e_G(V^*)|$.

\begin{fact}[Unit cost of SSR]
\label{fac:ssr}
Under SSR, for any graph $G$ and any target set $V^*$, the unit cost $\csf_u$ in \defref{perf} is a geometric RV with success probability $n^{e,*}_G/n$, i.e., $\csf_u^{\rm SSR} \sim \mathrm{geo}(n^{e,*}_G/n)$, and the expected unit cost is $c_u^{\rm SSR} = n/n^{e,*}_G$.
\end{fact}

\begin{IEEEproof}
SSR unit cost is the random number of independent Bernoulli trials until the first ``success'', i.e., the random star intersects the target set (equivalently, the random star center intersects the target set extended neighborhood).
\end{IEEEproof}

\begin{fact}[Unit cost of SSC]
\label{fac:ssc}
Under SSC, for any graph $G$ and any target set $V^*$, the expected unit cost $c_u^{\rm SSC}$ in \defref{perf} is $c_u^{\rm SSC} = (n+1)/(n^{e,*}_G + 1)$.  
\end{fact}

\begin{IEEEproof}
SSC with target set $V^*$ on a graph $G$ with extended neighborhood $N^e_G(V^*)$ is equivalent to sampling without replacement from an urn with $n$ balls, of which $n^{e,*}_G$ are marked, until a marked ball is drawn.  The expected number of samples is $(n+1)/(n^{e,*}_G+1)$ by \facref{ecnra}. 
\end{IEEEproof}

\facref{ssr} and \facref{ssc} show the expected number of SSR samples exceeds the expected number of SSC samples: $c_u^{\rm SSR} > c_u^{\rm SSC}$. 

We now adapt the previous two facts to the case where the initial graph is an ER random graph $\Gsf_0 = (V_0,\Esf_0)$ with parameters $(n,s)$, and where the expectation is with respect to both the graph and sampling distributions.  Define the RVs:
\begin{itemize}
\item $\nsf_t \equiv |\Vsf_t|$: order of graph $\Gsf_t$ (note $n_0 = n$);
\item $\nsf_t^* \equiv |\Vsf_t^*|$: number of vertices from the target set in $\Gsf_t$;
\item $\nsf_t^{e,*} \equiv |\Vsf_t^{e,*}|$ (where $\Vsf_t^{e,*} \equiv N_{\Gsf_t}^e(\Vsf_t^*)$):  number of vertices in extended neighborhood of target set in $\Gsf_t$.
\end{itemize} 

\begin{fact}
\label{fac:nsf0esER}
The random order of the extended neighborhood of the target set in the ER random graph $\Gsf_0$ is a shifted binomial RV:
\begin{equation}
\label{eq:nsf0esdisbn}
\nsf_0^{e,*} \sim n_0^* + \mathrm{bin}(n-n_0^*,1-(1-s)^{n_0^*}).
\end{equation}
Define $n_0^{e,*} \equiv \Ebb[\nsf_0^{e,*}]$, where 
\begin{eqnarray}
\label{eq:gpore}
\Ebb[\nsf_0^{e,*}] &=& n_0^* + (n-n_0^*)(1-(1-s)^{n_0^*}),\nonumber \\
\mathrm{var}(\nsf_0^{e,*}) &=& (n-n_0^*)(1-s)^{n_0^*}(1-(1-s)^{n_0^*}).
\end{eqnarray}
\end{fact}

\begin{IEEEproof}
Each $v \in V_0 \setminus V_0^*$ is connected to $V_0^*$ (independently of other vertices) if there exists an edge (or edges) from $v$ to $V_0^*$, which happens with probability $1 - (1-s)^{n_0^*}$.  
\end{IEEEproof}

\figref{gosm} (right) shows $n_0^{e,*}/n$, the expected fraction of vertices in the extended target set (\facref{nsf0esER}), vs.\ the edge probability $s$.  

The next result, from \cite{Lew1976}, is leveraged in \prpref{unicostssrsscer} below.

\begin{fact}[Bounds on inverse moments, \cite{Lew1976} (Eq.\ 3.1, p.\ 729)]
\label{fac:boundinvmom}
Let a random variable $\xsf$ have mean $\mu$, variance $\sigma^2$, and minimum support $x_{\rm min} > 0$ (i.e., $\Pbb(\xsf \geq x_{\rm min}) = 1$).  Then
\begin{equation}
\frac{1}{\mu} \leq \Ebb\left[\frac{1}{\xsf} \right] \leq \frac{1}{x_{\rm min}}
\frac{\sigma^2 + (\mu - x_{\rm min}) x_{\rm min}}{\sigma^2 + (\mu - x_{\rm min}) \mu}. \label{eq:bimgen}
\end{equation}
Specializing \eqnref{bimgen} to a binomial RV $\xsf \sim \mathrm{bin}(m,p)$, for $a > 0$:
\begin{equation}
\frac{1}{a+m p} \leq \Ebb\left[\frac{1}{a+\xsf} \right] \leq \frac{a+1-p}{a(a+1+(m-1)p)}.
\label{eq:bimbin}
\end{equation}
\end{fact}

\begin{proposition}[Unit cost of SSR and SSC for ER graph]
\label{prp:unicostssrsscer}
Fix the initial graph as an ER random graph $\Gsf_0$ with parameters $(n,s)$.  The expected unit cost under SSR and SSC has bounds
\begin{equation}
\underline{c}_u^{\rm SSR} \leq c_u^{\rm SSR} \leq \bar{c}_u^{\rm SSR}, ~~
\underline{c}_u^{\rm SSC} \leq c_u^{\rm SSC} \leq \bar{c}_u^{\rm SSC}.
\end{equation}
Let the target set $V_0^* \subseteq V_0$ have cardinality $n_0^* = |V_0^*|$:
\begin{eqnarray}
\underline{c}_u^{\rm SSR} & \equiv & \frac{n}{n - (1-s)^{n_0^*}(n-n_0^*)} \nonumber \\
\bar{c}_u^{\rm SSR} & \equiv & \frac{n((1-s)^{n_0^*} +n_0^*)}{n^*_0(n -(1-s)^{n_0^*}(n-1-n_0^*))} \nonumber \\
\underline{c}_u^{\rm SSC} & \equiv & \frac{n+1}{n+1-(1-s)^{n_0^*}(n-n_0^*)} \nonumber \\
\bar{c}_u^{\rm SSC} & \equiv & \frac{(n+1)((1-s)^{n_0^*} +1 +n_0^*)}{(n_0^*+1)(n+1-(1-s)^{n_0^*}(n-1- n_0^*))}. \label{eq:cuSSRcuSSClbubs2}
\end{eqnarray}
\end{proposition}

\begin{IEEEproof}
Using \facref{ssr}, the SSR bounds are derived by applying \eqnref{bimbin} in \facref{boundinvmom} to $n \Ebb[1/\nsf_0^{e,*}]$, for $\nsf_0^{e,*}$ in \eqnref{nsf0esdisbn} of \facref{nsf0esER}, i.e., with $a = n_0^*$, $m = n-n_0^*$, and $p = 1-(1-s)^{n_0^*}$.  Using \facref{ssc}, the SSC bounds are derived by applying \eqnref{bimbin} in \facref{boundinvmom} to $(n+1)\Ebb[1/(\nsf_0^{e,*}+1)]$, i.e., with $a = n_0^*+1$, $m = n-n_0^*$, and $p = 1-(1-s)^{n_0^*}$.  
\end{IEEEproof}

The following approximation and derivation was suggested by one of the Anonymous Reviewers; c.f.\ \figref{ucadr}.

\begin{proposition}[Approximate unit cost of SSR for ER graph]
\label{prp:approxunicostssr}
Fix the initial graph as an ER random graph $\Gsf_0$ with parameters $(n,s)$.  The approximate expected unit cost under SSR is
\begin{equation}
c_u^{\rm SSR} \approx \tilde{c}_u^{\rm SSR} \equiv \frac{n}{n_0^*(1+s(n-1))}.
\end{equation}
This approximation equals the lower bound $\underline{c}_u^{\rm SSR}$ for $n_0^* = 1$.  
\end{proposition}

\begin{IEEEproof}
Recall $\dsf,\msf,\ssf$ from \secref{ergp}.  Ignoring {\em duplicates}, the expected number of {\em vertices} observed after $c_u^{\rm SSR}$ samples is $(1+\Ebb[\dsf]) c_u^{\rm SSR}$, for $\dsf = 2\msf/n = \ssf(n-1)$ the average degree.  As the fraction of vertices in the target set is $n_0^*/n$, it follows that the expected number of {\em target set vertices} observed after $c_u^{\rm SSR}$ samples is $(1+s(n-1)) c_u^{\rm SSR} n_0^*/n$.  As the sampling terminates after the first observed target set vertex is found, an approximation of $c_u^{\rm SSR}$, denoted $\tilde{c}_u^{\rm SSR}$, is the solution of $(1+s(n-1)) \tilde{c}_u^{\rm SSR} n_0^*/n = 1$.
\end{IEEEproof}


\subsection{Unit cost analysis for SSS}
\label{sec:ucs}

Let $(\ysf,\zsf)$ be a pair of continuous RVs, with $\zsf \neq 0$ almost surely, means $(\mu_{\ysf},\mu_{\zsf})$, variances $(\sigma^2_{\ysf},\sigma^2_{\zsf})$, and covariance $\mathrm{cov}(\ysf,\zsf)$.  Consider the expectation of their ratio, i.e., $\Ebb \left[ \frac{\ysf}{\zsf} \right]$.  The following result is found in \cite{KemVli2000} and \cite{Sel2017}.

\begin{proposition}[\cite{KemVli2000}, \cite{Sel2017}]
\label{prp:prestratio}
The second order Taylor series approximation of $\Ebb \left[ \frac{\ysf}{\zsf} \right]$ around $(\mu_{\ysf},\mu_{\zsf})$ is
\begin{equation}
\Ebb \left[ \frac{\ysf}{\zsf} \right] \approx \frac{\mu_{\ysf}}{\mu_{\zsf}} + \frac{\sigma_{\zsf}^2 \mu_{\ysf}}{\mu_{\zsf}^3} - \frac{\mathrm{cov}(\ysf,\zsf)}{\mu_{\zsf}^2}.
\end{equation}
\end{proposition}

In particular, the error associated with a first-order Taylor series approximation, defined as
\begin{equation}
\label{eq:taylorerrordef1}
\epsilon_1 \equiv \left| \Ebb \left[ \frac{\ysf}{\zsf} \right] - \frac{\mu_{\ysf}}{\mu_{\zsf}} \right|,
\end{equation}
is approximated as
\begin{equation}
\label{eq:taylorerrordef2}
\tilde{\epsilon}_1 \equiv \left| \frac{\sigma_{\zsf}^2 \mu_{\ysf}}{\mu_{\zsf}^3} - \frac{\mathrm{cov}(\ysf,\zsf)}{\mu_{\zsf}^2} \right|.
\end{equation}
As clarification, the first-order Taylor series approximation has an approximation error denoted $\epsilon_1$; this approximation error is itself approximated as $\tilde{\epsilon}_1$.  This fact is used in \thmref{lfeo} below.

Let the initial graph be an ER random graph $\Gsf_0 = (V_0,\Esf_0)$ with parameters $(n,s)$.  Recall a star will hit the target set if its star center is in the extended neighborhood of the target.  The (random) probability that star $t+1$ hits the target set is 
\begin{equation}
\psf_{t+1}^{\rm SSS} \equiv \frac{\nsf^{e,*}_t}{\nsf_t}.
\end{equation}
Define events $(\Emc_t, t \in \Zbb^+)$, with $\Emc_0$ trivial, and $\Emc_t \equiv \left\{ \vsf_t \not\in \Vsf^{e,*}_{t-1}) \right\}$ the event that star $t$ misses the target set.  Next, define events $(\bar{\Emc}_t, t \in \Zbb^+)$, with $\bar{\Emc}_0$ trivial, and 
\begin{equation}
\bar{\Emc}_t \equiv \bigcap_{t' \in [t]} \Emc_{t'} = \left\{ \vsf_{t'} \not\in \Vsf^{e,*}_{t'-1}, ~ \forall t' \in [t] \right\}.
\end{equation}
Thus, $\bar{\Emc}_t$ is the event that the stars of the first $t$ samples have each missed the target set.  Observe that, conditioned on $\bar{\Emc}_t$, the target set in graph $\Gsf_t$ is identical to the same set in the initial graph $\Gsf_0$, i.e., $\Vsf^*_t | \bar{\Emc}_t = V_0^*$, although the degrees of vertices in $V_0^*$ may have decreased due to sampling.  

The expected probability of hitting the target with star $t+1$, conditioned on missing the target set in the first $t$ samples, is:
\begin{equation}
\Ebb[\psf_{t+1}^{\rm SSS} | \bar{\Emc}_t] = \Ebb \left[ \left. \frac{\nsf^{e,*}_t}{\nsf_t} \right| \bar{\Emc}_t \right].
\end{equation}
Via \eqnref{taylorerrordef1} and \eqnref{taylorerrordef2}, the error in approximating this expected conditional probability by its ratio of expectations, i.e., 
\begin{equation}
\tilde{p}_{t+1}^{\rm SSS} \equiv \frac{\Ebb[\nsf^{e,*}_t| \bar{\Emc}_t]}{\Ebb[\nsf_t| \bar{\Emc}_t]}, 
\label{eq:approxpt}
\end{equation}
is approximately
\begin{equation}
\label{eq:nfoc}
\tilde{\epsilon}_{n,t} \equiv \left| \frac{\mathrm{var}(\nsf_t| \bar{\Emc}_t) \Ebb[\nsf^{e,*}_t| \bar{\Emc}_t]}{\Ebb[\nsf_t| \bar{\Emc}_t]^3} - \frac{\mathrm{cov}(\nsf^{e,*}_t,\nsf_t| \bar{\Emc}_t)}{\Ebb[\nsf_t| \bar{\Emc}_t]^2} \right|.
\end{equation}
This approximate {\em conditional} hitting probability and its approximation error are expressed in terms of the three fundamental parameters $(n,n_0^*,s)$ and the sample index $t$ in \thmref{lfeo}.

\begin{theorem}
\label{thm:lfeo}
The approximate probability of hitting the target set in sample $t+1$ under SSS, conditioned on missing it in the first $t$ samples, is (for $n_0^{e,*} \equiv \Ebb[\nsf_0^{e,*}]$ in \facref{nsf0esER} \eqnref{gpore}):
\begin{equation}
\label{eq:paucl}
\tilde{p}_{t+1}^{\rm SSS} = \frac{(n_0^{e,*}-n_0^*) (1-s)^t +n_0^*}{n (1-s)^t - \frac{1-s}{s} (1 - (1-s)^t)}.
\end{equation}
Viewing $\tilde{p}_{t+1}^{\rm SSS}$ as a continuous function of $t$, $\tilde{p}_{t+1}^{\rm SSS}$ is convex increasing in $t$ over $t \in [0,t^{(2)}_{\tilde{p}})$, where
\begin{equation}
\label{eq:t2tildep}
t^{(2)}_{\tilde{p}} \equiv \frac{\log\left(n \frac{s}{1-s}+1 \right)}{\log(1/(1-s))}.
\end{equation}
Moreover, $\tilde{p}_1^{\rm SSS}$ (hitting in the first sample) equals the exact value $p_1^{\rm SSS} \equiv n^{e,*}_0/n$, and $\tilde{p}_{t+1}^{\rm SSS} = 1$ at $t^{(1)}_{\tilde{p}} \equiv$
\begin{equation}
\label{eq:t1tildep}
\frac{\log\left( s(n-\Ebb[\nsf_0^{e,*}]+n^*_0) + (1-s) \right) -\log\left((1-s) + n^*_0 s\right)} {\log(1/(1-s))}.
\end{equation}
It may be shown that $t^{(1)}_{\tilde{p}}  < t^{(1)}_{\tilde{p}}$. The approximate error in this approximation is $\tilde{\epsilon}_{n,t}=$
\begin{eqnarray}
& & \left( n (1 - (1-s)^t) + \frac{(1-s)(1 + (1-s)(1-(1-s)^t))}{s(1+(1-s))}\right) \nonumber \\
& & \times \frac{(1-s)^t((n_0^{e,*} -n_0^*)(1-s)^t +n_0^*)}{\left(n (1-s)^t - \frac{1-s}{s} (1 - (1-s)^t) \right)^3}. \label{eq:epsilonntucl}
\end{eqnarray}
The approximate error in this approximation is asymptotically negligible in $n$:
\begin{equation}
\lim_{n \uparrow \infty} \tilde{\epsilon}_{n,t} = 0, ~ \forall t \in \Nbb, s \in (0,1). \label{eq:asymperrorucl}
\end{equation}
and in fact $\tilde{\epsilon}_{n,t} = O(n_0^{e,*}/n^2)$.
\end{theorem}

\begin{figure}[ht]
\centering
\includegraphics[width=\columnwidth]{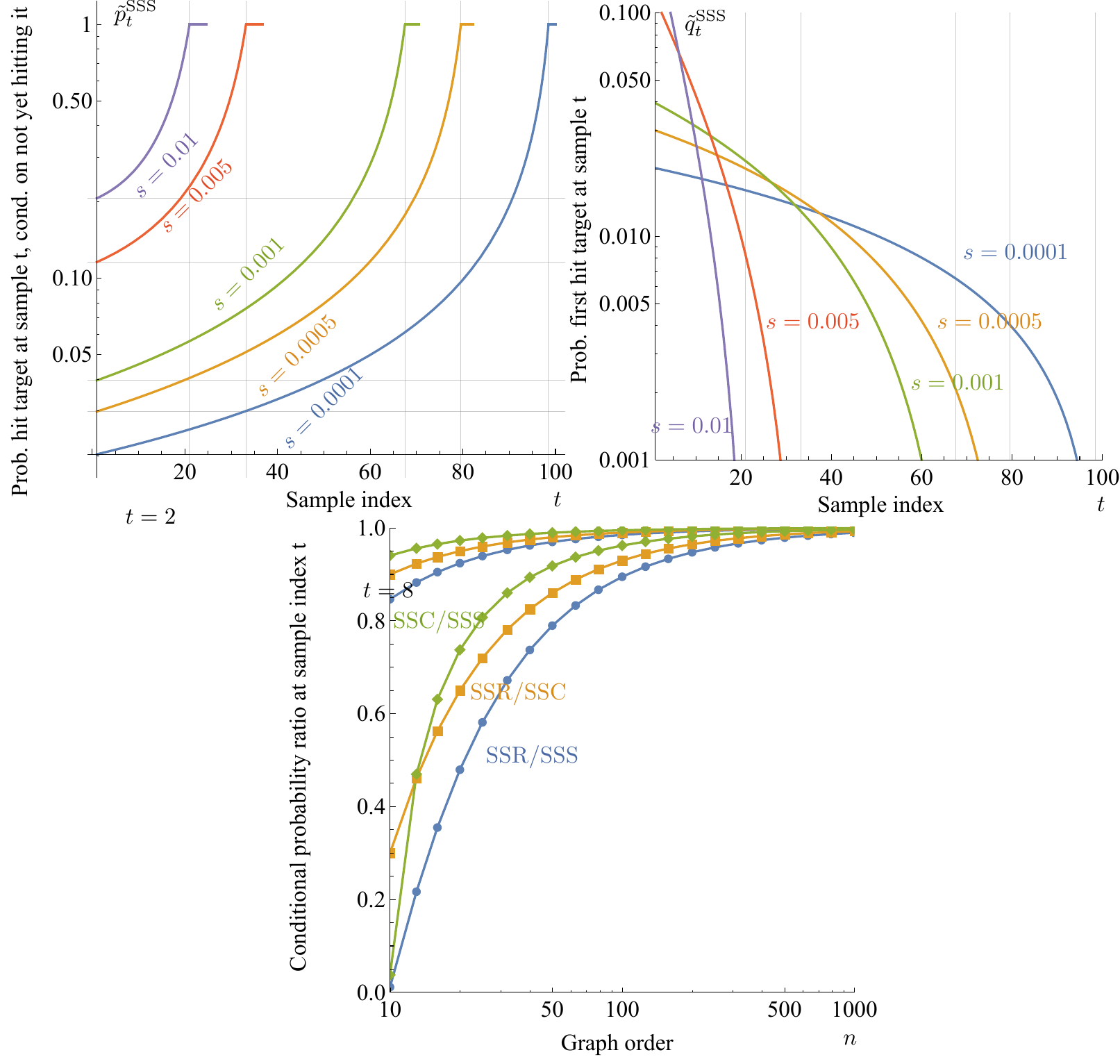}
\caption{
{\em Top left:} \thmref{lfeo}: approximate {\em conditional} probability of first hitting the target in an ER graph with parameters $(n=100,s)$ at sample $t$, conditioned on not yet hitting it, under SSS, i.e., $\tilde{p}_t^{\rm SSS}$ from \eqnref{paucl}, vs.\ $t$, for $n_0^* = 2$ and $s \in \{0.0001,0.0005,0.001,0.005,0.01\}$.  Horizontal lines show $p_1^{\rm SSS} \equiv n^{e,*}_0/n$ and vertical lines show $t^{(1)}_{\tilde{p}}$ from \eqnref{t1tildep}.
{\em Top right:} \thmref{appuncprobhittarsetSSS}: approximate {\em unconditional} probability of first hitting the target in an ER graph with parameters $(n=100,s)$ at sample $t$ under SSS, i.e., $\tilde{q}_t^{\rm SSS}$ from \eqnref{tildeqtsss}, vs.\ $t$, for $n_0^* = 2$ and $s \in \{0.0001,0.0005,0.001,0.005,0.01\}$.  Vertical lines show $t^{(1)}_{\tilde{p}}$ from \eqnref{t1tildep}.
{\em Bottom:} \prpref{epgs}: three conditional probability ratios vs.\ graph order $n$, for $t \in \{2,8\}$ ($n_0^* = 2$, $s(n) = 1/n$).}
\label{fig:lfeo}
\end{figure}

The proof is found in \appref{lfeo}.  \figref{lfeo} (top left) shows $\tilde{p}_t^{\rm SSS}$ from \eqnref{paucl} vs.\ the sample index $t$.  Note the curves are convex increasing even on a logarithmic scale, i.e., $\tilde{p}_t^{\rm SSS}$ is log-convex. 

The approximate {\em unconditional} hitting probability $\tilde{q}_{t}^{\rm SSS}$ and the corresponding approximate expected unit cost under SSS for an ER graph $\tilde{c}_u^{\rm SSS}$ are expressed in terms of the three fundamental parameters $(n,n_0^*,s)$ in \thmref{appuncprobhittarsetSSS}.  

Let the RV $\tilde{\csf}_u^{\rm SSS}$ have distribution $\tilde{q}^{\rm SSS} \equiv (\tilde{q}_t^{\rm SSS}, t \in [t_{\tilde{p}}^{(1)}])$, so that the approximate expected unit cost is the expectation of this RV with respect to that distribution: $\tilde{c}_u^{\rm SSS} \equiv \Ebb[\tilde{\csf}_u^{\rm SSS}]$.

\begin{theorem}
\label{thm:appuncprobhittarsetSSS}
The approximate {\em unconditional} probability of first hitting the target set in sample $t$ under SSS is
\begin{equation}
\label{eq:tildeqtsss}
\tilde{q}_{t}^{\rm SSS} \equiv \tilde{p}_{t}^{\rm SSS} \prod_{u \in [t-1]} (1 - \tilde{p}_{u}^{\rm SSS}), 
\end{equation}
for $t \in [ \lfloor t_{\tilde{p}}^{(1)} \rfloor ]$, with $t_{\tilde{p}}^{(1)}$ defined in \eqnref{t1tildep}.  The associated approximate {\em expected} unit cost under SSS is
\begin{equation}
\tilde{c}_u^{\rm SSS} \equiv \sum_{t \in [t_{\tilde{p}}^{(1)}]} \prod_{u \in [t]} (1 - \tilde{p}_{u}^{\rm SSS}).
\label{eq:approxexpunitcostundersss}
\end{equation}
\end{theorem}

The proof is found in \appref{appuncprobhittarsetSSS}.  \figref{lfeo} (top right) shows $\tilde{q}_t^{\rm SSS}$ from \eqnref{tildeqtsss} vs.\ the sample index $t$.  As $s$ increases the distribution support decreases and the distribution slope decreases.


The last result in this section, \prpref{epgs}, considers the parameter regime where the ER edge probability $s$ is $O(1/n)$.  Let 
\begin{equation}
p_t^{\rm SSR} = \frac{n_0^{e,*}}{n}, ~ p_t^{\rm SSC} = \frac{n_0^{e,*}}{n-t+1}
\end{equation}
be the conditional probabilities of hitting the target set for the first time on sample $t$ under SSR and SSC, respectively.

\begin{proposition}
\label{prp:epgs}
Set the ER edge probability $s(n)=c/n$ (with $c>0$), and fix $t$ and $n_0^*$.  The {\em conditional} probability of hitting the target set in sample $t$ (conditioned on not yet having hit it) is, asymptotically as $n \uparrow \infty$, equal for all three SS variants:
\begin{equation}
\frac{\tilde{p}_{t}^{\rm SSS}}{p_{t}^{\rm SSR}} \to 1, \; 
\frac{p_{t}^{\rm SSR}}{p_{t}^{\rm SSC}} \to 1, \; 
\frac{\tilde{p}_{t}^{\rm SSS}}{p_{t}^{\rm SSC}} \to 1. 
\end{equation}
\end{proposition}

The proof is found in \appref{epgs}.  \figref{lfeo} (bottom) shows all three conditional probability ratios in \prpref{epgs} vs.\ the graph order $n$, for $n_0^* = 2$ and $s(n) = c/n$ with $c=1$ for $t \in \{2,8\}$.  The convergence of all three ratios to $1$ as $n \uparrow \infty$ is evident.


\subsection{Numerical results}
\label{sec:ucnr}

\begin{figure}[ht!]
\centering
\includegraphics[width=\columnwidth]{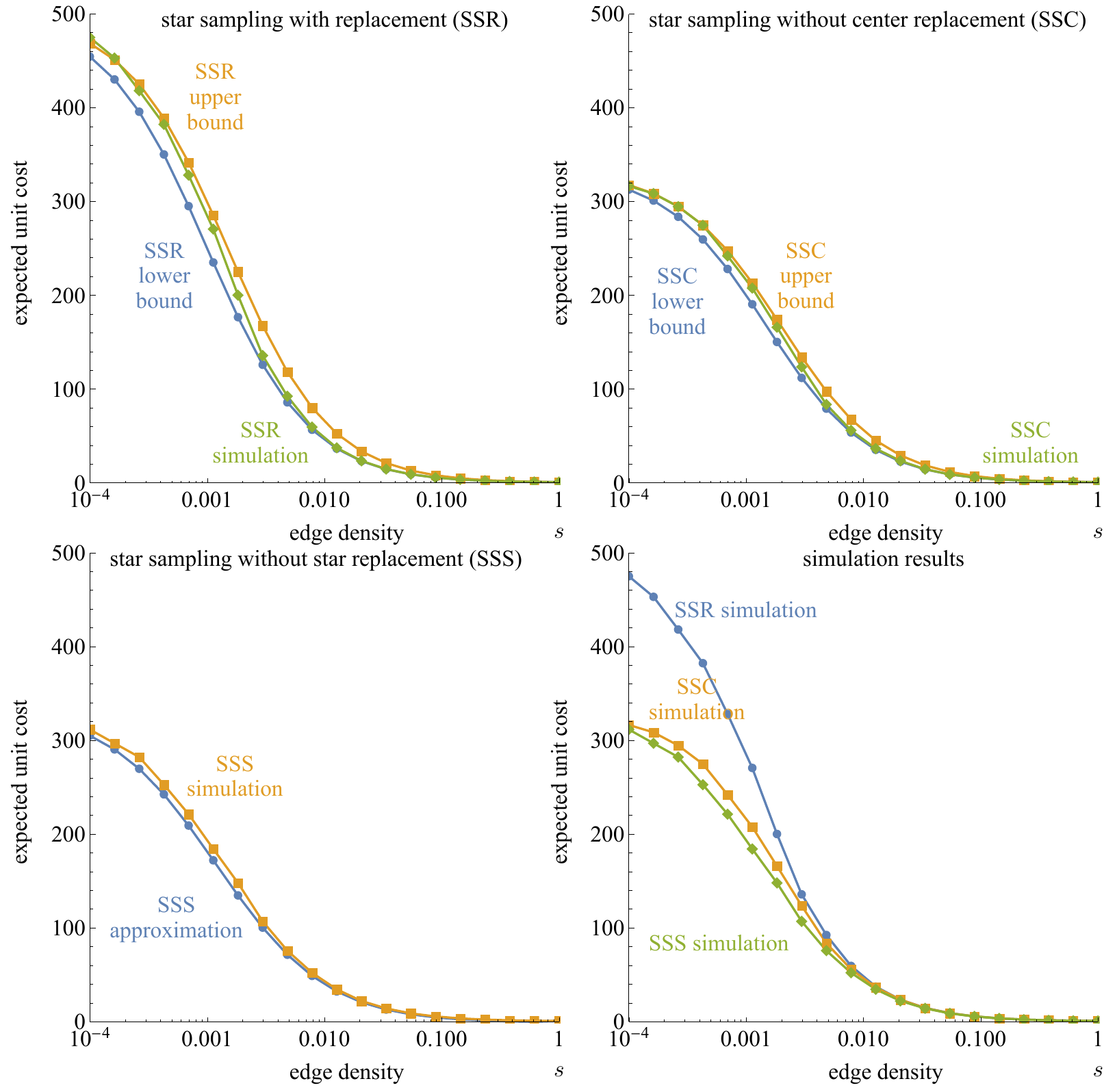}
\caption{Numerical and simulation results for expected {\em unit} cost on ER random graphs with order $n=1,000$ and target size $n_0^* = 2$ vs.\ edge density $s$ (on log scale).  {\em Top left:} SSR bounds and simulation.  {\em Top right:} SSC bounds and simulation.  {\em Bottom left:} SSC approximation and simulation.  {\em Bottom right:} simulation results for SSR, SSC, SSS.}
\label{fig:ucnr}
\end{figure}

The four plots in \figref{ucnr} show numerical and simulation results for the expected {\em unit} cost to find $v\in V^*$ (with $n_0^* = 2$) under SSR, SSC, and SSS on ER random graphs with order $n=1,000$ vs.\ the edge density $s$ (on a log scale).  Three conclusions are evident from the plots. First, the bounds on SSR (\prpref{unicostssrsscer}, top left plot) and SSC (\prpref{unicostssrsscer}, top right plot) are ``tight'' and the approximation for SSS (\thmref{lfeo}, bottom left plot) is ``accurate''.  Second, the average unit cost ordering on ER graphs is $c_u^{\rm SSR} \geq c_u^{\rm SSC} \geq c_u^{\rm SSS}$ (bottom right plot).  Third, there is a notable difference in unit cost between SSR and SSC/SSS for ``small'' $s$, but no appreciable difference among the three sampling schemes for ``large'' $s$ (bottom right plot).

\begin{figure}[ht!]
\centering
\includegraphics[width=\columnwidth]{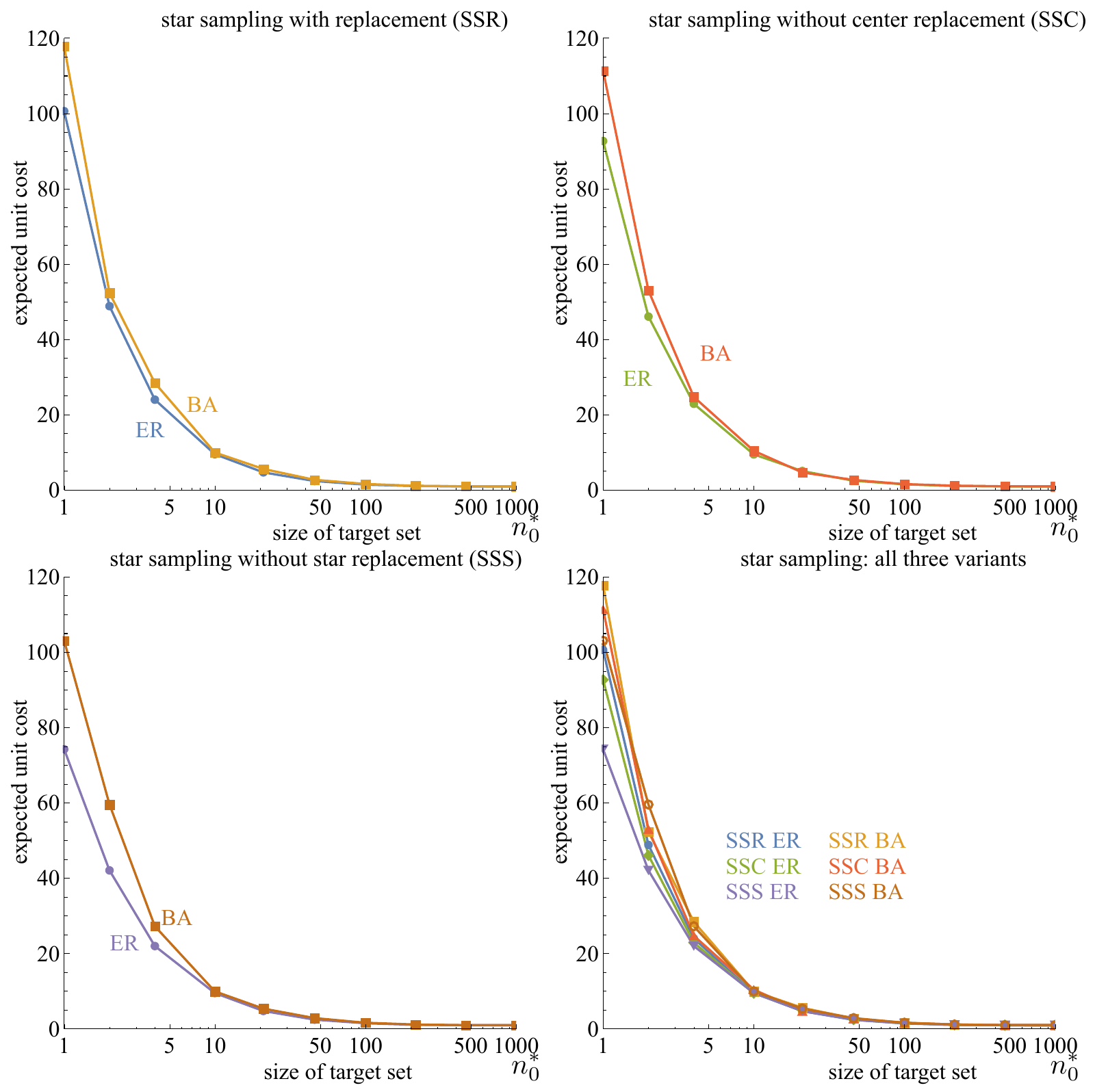}
\caption{Simulation results for expected {\em unit} cost on ER and BA random graphs with order $n=1,000$ and $s = 0.01$ vs.\ target set size $n_0^*$ (on log scale).  {\em Top left:} SSR.  {\em Top right:} SSC.  {\em Bottom left:} SSC.  {\em Bottom right:} all three star sampling types (SSR, SSC, SSS).}
\label{fig:ucerba}
\end{figure}

As results have been developed for the special case of ER graphs, it is natural to investigate the impact of the graph structure on the expected unit cost.  Towards that end, consider the case of Barab\'{a}si-Albert (BA) random graphs, in which (roughly) each of the $n$ vertices is ``added'' sequentially, and each ``new'' vertex adds $m$ edges, which are attached to the existing edges using preferential attachment, i.e., with probability on each vertex proportional to the vertex degree.  As a BA graph with parameters $(n,m)$ has approximately $n m$ edges, such a graph will have similar order and size to an ER graph with parameters $(n,s)$ provided $n m \approx \binom{n}{2} s$, i.e., $2 m \approx (n-1) s$.  The results below use $n = 1,000$, $s = 0.01$, and $m = 5$.  The four plots in \figref{ucerba} show simulation results (each point the average over $1,000$ trials) for the expected {\em unit} cost to find $v\in V^*$ under SSR, SSC, and SSS on both ER and BA random graphs vs.\ the target set size $n_0^*$ (on a log scale).  It is evident from the plots that there is only a small difference in expected unit cost for BA and ER graphs and the BA vs.\ ER unit cost difference decreases in $n_0^*$.  Therefore, although analytical results and approximations are derived under the assumption of an ER graph, they  appear to hold reasonably well for at least one other ``class'' of random graphs.

\begin{figure}[ht!]
\centering
\includegraphics[width=\columnwidth]{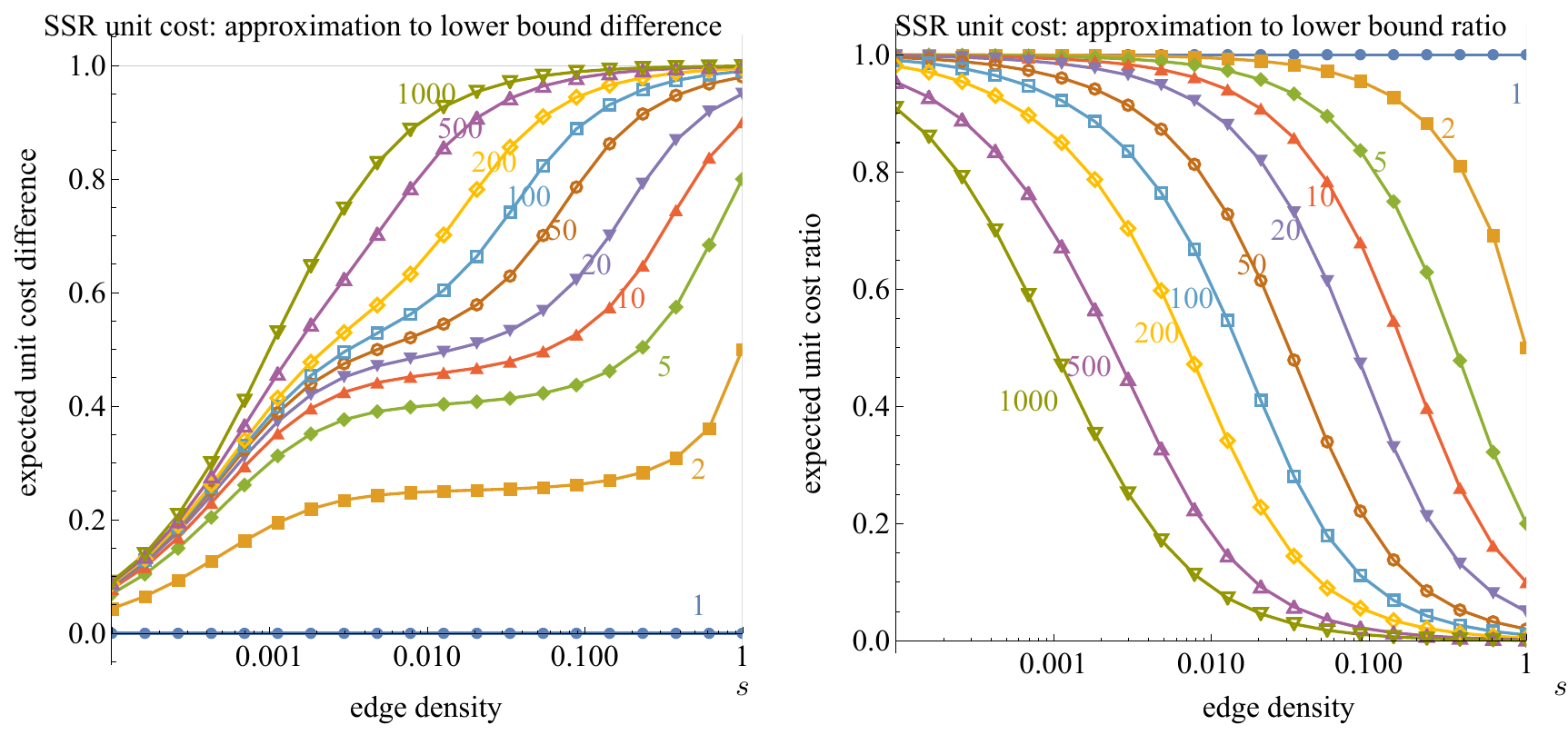}
\caption{Numerical results for SSR expected unit cost approximation $\tilde{c}_u^{\rm SSR}$ from \prpref{approxunicostssr} on ER random graphs with order $n=1,000$ and variable target size $n_0^*$ (labeled curves $\{1,2,\ldots,500,1000\}$) vs.\ edge density $s$ (on log scale).  {\em Left:} expected cost {\em difference} (lower bound minus approximation).  {\em Right:} expected cost {\em ratio} (approximation over lower bound).}
\label{fig:ucadr}
\end{figure}

The two plots in \figref{ucadr} show numerical results for the SSR expected unit cost approximation $\tilde{c}_u^{\rm SSR}$ from \prpref{approxunicostssr} (relative to the lower bound $\underline{c}_u^{\rm SSR}$ from \prpref{unicostssrsscer}) on ER random graphs with order $n=1,000$ vs.\ the edge density $s$ (on a log scale), for various values of $n_0^*$.  The {\em left} plot shows the {\em difference} of the lower bound minus the approximation ($\underline{c}_u^{\rm SSR}-\tilde{c}_u^{\rm SSR}$), and the {\em right} plot shows the {\em ratio} of the approximation over the lower bound ($\tilde{c}_u^{\rm SSR}/\underline{c}_u^{\rm SSR}$).  The plots suggest $i)$ the approximation is always {\em lower than} but within $1$ of the lower bound cost for all values of $(n_0^*,s)$ and $ii)$ the approximation is highly accurate, both as a difference and as a ratio, for the typical case of small $n_0^*$ and small $s$.

\section{Linear Cost Model}
\label{sec:s_cost}

The expected linear cost (\defref{perf2}) of SSR on an arbitrary graph is given in \secref{lcagssr}.  Approximate expected linear costs of SSR, SSC, and SSS on an ER random graph are given in \secref{lcerall}.  Numerical results are presented in \secref{lcnr}.  

\subsection{Arbitrary graph}
\label{sec:lcagssr}

Let $G = (V,E)$ be an arbitrary graph with order $|V| = n$, and let $V^* \subseteq V$ be an arbitrary target set. \facref{lcagssr} below is the linear cost analog of the unit cost result for SSR in \facref{ssr}.  

We first develop some necessary notation, extending that introduced in \secref{notation}.  Let $V^{e,*}_G = N_G^e(V^*)$ denote the extended neighborhood of $V^*$ (with order $n^{e,*}_G \equiv |V^{e,*}_G|$), and let $\bar{V}^{e,*}_G \equiv V \setminus V^{e,*}_G$ denote its complement. Define $V^{e,*}_G(k) \equiv \{ v \in V^{e,*}_G : d_G(v) = k\}$ and $\bar{V}^{e,*}_G(k) \equiv \{ v \in \bar{V}^{e,*}_G : d_G(v) = k\}$ as the subsets of vertices in $V^{e,*}_G,\bar{V}^{e,*}_G$, respectively, of degree $k$.  The conditional degree distributions in $V^{e,*}_G,\bar{V}^{e,*}_G$ are denoted $w^{e,*}_G,\bar{w}^{e,*}_G$, respectively, with components
\begin{equation}
w^{e,*}_G(k) \equiv \frac{|V^{e,*}_G(k)|}{|V^{e,*}_G|}, ~
\bar{w}^{e,*}_G(k) \equiv \frac{|\bar{V}^{e,*}_G(k)|}{|\bar{V}^{e,*}_G|}.
\end{equation}
Finally, the average degrees in $V^{e,*}_G,\bar{V}^{e,*}_G$, respectively, are
\begin{equation}
d^{e,*}_G = \sum_{k \in D^{e,*}_G} k w^{e,*}_G(k), ~
\bar{d}^{e,*}_G = \sum_{k \in \bar{D}^{e,*}_G} k \bar{w}^{e,*}_G(k).
\end{equation}

\begin{fact}[Linear cost of SSR]
\label{fac:lcagssr}
Under SSR, for any graph $G$ and any target set $V^*$, the expected linear cost $c_l$ in \eqnref{perfdef4} is 
\begin{equation}
\label{eq:clssrag}
c_l^{\rm SSR}(G,V^*) = \bar{d}^{e,*}_G \left( \frac{n}{n^{e,*}_G} - 1 \right) + d^{e,*}_G.
\end{equation}
\end{fact}

The proof is found in \appref{lcagssr}.

Despite the success in deriving the previous result for SSR, we are of the impression that a simple exact expression for the expected linear cost under SSC on an arbitrary graph is not available.  In contrast to the {\em unit} cost case, for which the SSC expected unit cost is given by \facref{ssc}, the linear cost case appears to be much more difficult to analyze, as it requires a representation of the evolution under SSC of the (effectively arbitrary) degree distribution $w_G$ of the initial graph $G$.  This evolution does not appear to be sufficiently tractable to yield ``closed-form'' results like those in \secref{ucrc}.  Given this difficulty, we turn to approximations for the case of an ER random graph.

\subsection{ER random graph}
\label{sec:lcerall}

Let the initial graph be an ER random graph $\Gsf_0 = (V_0,\Esf_0)$ with parameters $(n,s)$.  Expectations below are taken with respect to the graph and sampling distributions. The starting point for all three sampling paradigms is to express the expected linear cost in terms of the conditional expectations of the extended degrees of each sampled vertex, conditioned on the unit cost, and to {\em approximate} the per-sample extended degree as {\em independent} of the unit cost:
\begin{eqnarray}
c_l 
& \equiv & \Ebb[\csf_l] 
= \Ebb\left[ \sum_{t \in [c_{\rm max}]} \dsf^e_t \mathbf{1}(t \leq \csf_u) \right] \label{eq:eclapproxuc} \\
&=& \sum_{t \in [c_{\rm max}]} \Ebb[\dsf^e_t \mathbf{1}(t \leq \csf_u)] 
\approx \sum_{t \in [c_{\rm max}]} \Ebb[\dsf^e_t] \Pbb(t \leq \csf_u). \nonumber 
\end{eqnarray}
Here, $\dsf^e_t$ is the extended degree of sample $t$, $\csf_u$ is the unit cost, i.e., the random number of samples required for the star center to hit the extended target set, and $c_{\rm max}$ is an upper bound, possibly infinite, on $\csf_u$.  The extended degree is {\em not} independent of the number of samples, as the target set in $\Gsf_t$ will have a distinct degree distribution from its complement, conditioned on the previous $t-1$ samples missing the target.  Nonetheless, the approximation is useful in that it facilitates analysis and our numerical investigations support our claim that the approximation is accurate over a wide array of parameter values. Below, we adapt the above approximation to give the approximate expected linear cost {\em conditioned} on the random extended target set, $\nsf_0^{e,*}$, i.e., 
\begin{equation}
\label{eq:eclapprox}
\Ebb[\tilde{\csf}_l|\nsf_0^{e,*}] \approx \tilde{\Ebb}[\tilde{\csf}_l|\nsf_0^{e,*}] \equiv \sum_{t \in [c_{\rm max}(\nsf_0^{e,*})]} \Ebb[\dsf^e_t] \Pbb(t \leq \csf_u | \nsf_0^{e,*}).
\end{equation}
The additional approximation in \eqnref{eclapprox}, beyond the one in \eqnref{eclapproxuc}, is that $\dsf^e_t$ is independent of $\nsf_0^{e,*}$, i.e., $\Ebb[\dsf^e_t|\nsf_0^{e,*}] = \Ebb[\dsf^e_t]$.

\subsubsection{SSR}

In the case of SSR, the approximation in \eqnref{eclapprox} leads to a particularly simple expression on account of the fact that $(\dsf^e_t)$ are IID, i.e., $\Ebb[\dsf^e_t]$ does not depend upon $t$, and as such Wald's identity for the expected value of the sum of a random number of IID RVs applies:
\begin{eqnarray}
\tilde{\Ebb}[\tilde{\csf}_l^{\rm SSR}|\nsf_0^{e,*}]
& = & \Ebb[\dsf^e_1] \sum_{t \in [c_{\rm max}^{\rm SSR}(\nsf_0^{e,*})]} \Pbb(t \leq \csf_u|\nsf_0^{e,*}) \\
& = & \Ebb[\dsf^e_1] \Ebb[\csf_u|\nsf_0^{e,*}] = (1 + (n-1)s) \frac{n}{\nsf_0^{e,*}}. \nonumber 
\end{eqnarray}
Here, $\Ebb[\dsf^e_1] = 1 + (n-1)s$ is the expected extended degree of a random vertex in an ER random graph, and $\Ebb[\csf_u|\nsf_0^{e,*}] = n/\nsf_0^{e,*}$ is the conditional expected unit cost of SSR.  Taking expectation with respect to $\nsf_0^{e,*}$ yields the following.

\begin{proposition}[SSR linear cost for ER graph.]
\label{prp:lcssrer}
Fix the initial graph as an ER random graph $\Gsf_0$ with parameters $(n,s)$, and the extended target set cardinality $\nsf_0^{e,*}$.  The approximate expected linear cost under SSR, $\tilde{c}_l^{\rm SSR}$, is
\begin{equation}
\tilde{c}_l^{\rm SSR} \equiv \Ebb[\Ebb[\tilde{\csf}_l^{\rm SSR}|\nsf_0^{e,*}]] = (1 + (n-1)s) c_u^{\rm SSR},
\end{equation}
and the SSR expected unit cost, $c_u^{\rm SSR}$, has bounds in \prpref{unicostssrsscer}.
\end{proposition}

\subsubsection{SSC}

Leveraging the approximation \eqnref{eclapprox} for SSC requires the expected extended degree of the random star center selected in sample $t$ and the hitting probability for sample $t$. 

\begin{proposition}[SSC linear cost for ER graph.]
\label{prp:lcsscer}
The approximate expected linear cost under SSC for an ER graph, conditioned on the random extended target set order $\nsf^{e,*}_0$, is 
\begin{equation}
\label{eq:aelcsscercetso}
\Ebb[\tilde{\csf}_l^{\rm SSC} | \nsf^{e,*}_0] = \sum_{t=1}^{n - \nsf^{e,*}_0+1} (1+(n-t)s) \prod_{u=1}^{t-1} \left(1 - \frac{\nsf^{e,*}_0}{n-u+1} \right).
\end{equation}
\end{proposition} 

The proof is found in \appref{lcsscer}.

\subsubsection{SSS} 

As with SSC, leveraging the approximation \eqnref{eclapprox} for SSS requires the expected extended degree of the random star center selected in sample $t$ and the approximate conditional hitting probability for sample $t$. 

\begin{proposition}[SSS linear cost for ER graph.]
\label{prp:lcssser}
The approximate expected linear cost under SSS for an ER graph, conditioned on the random extended target set order $\nsf^{e,*}_0$, is 
\begin{equation}
\label{eq:aelcsssercetso}
\Ebb[\tilde{\csf}_l^{\rm SSS} | \nsf^{e,*}_0] 
= \sum_{t=1}^{t^{(1)}_{\tilde{p}}} ((n-1)s + 1)(1-s)^{t-1} \prod_{u=1}^{t-1} \left(1 - \tilde{p}_{u}^{\rm SSS} \right),
\end{equation}
where $t^{(1)}_{\tilde{p}}$ (c.f.\ \eqnref{t1tildep}) and $\tilde{p}_t^{\rm SSS}$ (c.f.\ \eqnref{paucl}) are defined in \thmref{lfeo}.
\end{proposition} 

The proof is found in \appref{lcssser}.

\subsection{Numerical results}
\label{sec:lcnr}

\begin{figure}[ht!]
\centering
\includegraphics[width=\columnwidth]{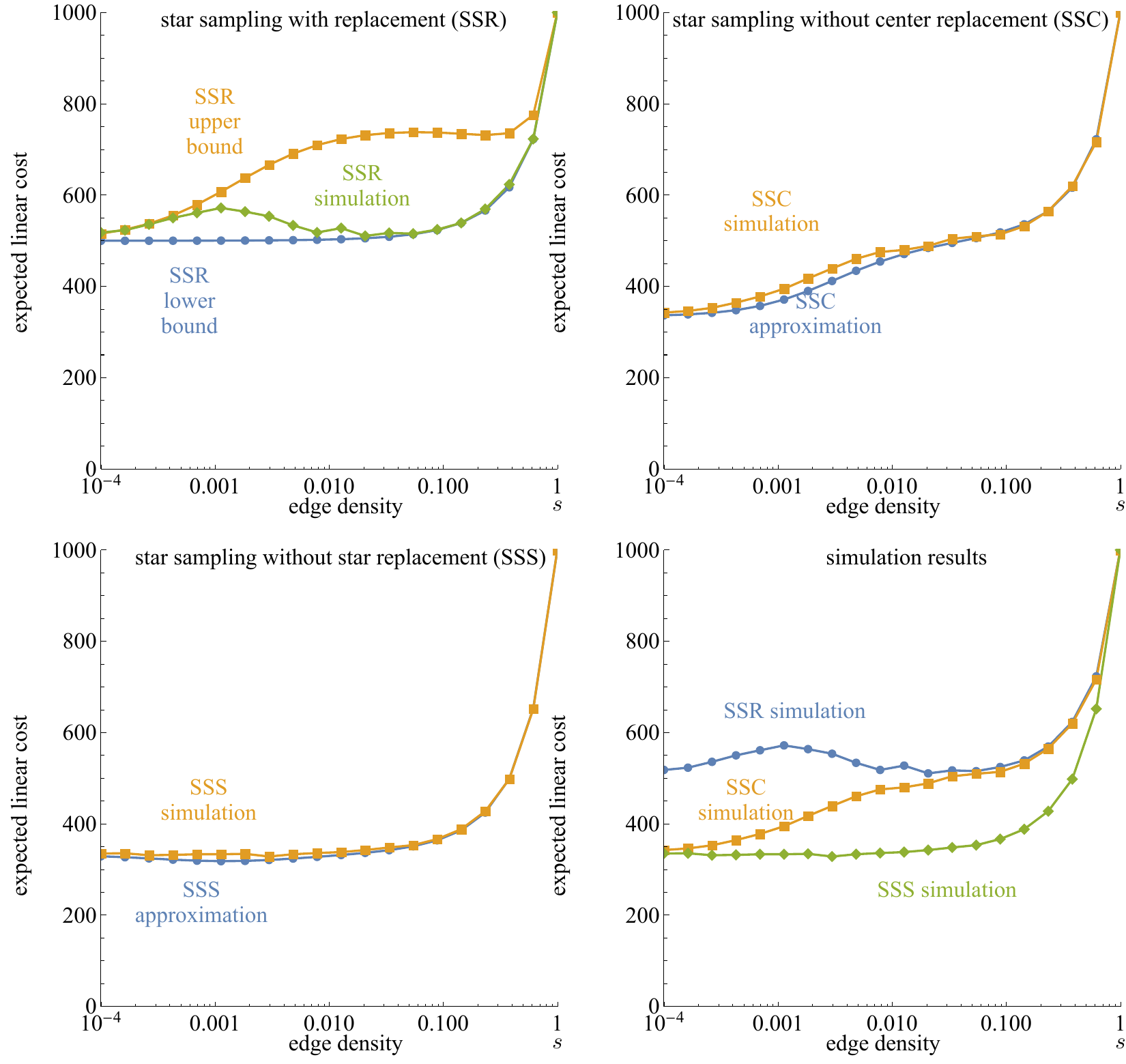}
\caption{Numerical and simulation results for expected {\em linear} cost on ER random graphs with order $n=1,000$ and target size $n_0^* = 2$ vs.\ edge density $s$ (on log scale).  {\em Top left:} SSR bounds and simulation.  {\em Top right:} SSC approximation and simulation.  {\em Bottom left:} SSC approximation and simulation.  {\em Bottom right:} simulation results for SSR, SSC, SSS.}
\label{fig:lcnr}
\end{figure}

The four plots in \figref{lcnr} show numerical and simulation results for the expected {\em linear} cost to find $v \in V^*$ (with $n_0^* = 2$) under SSR, SSC, and SSS on ER random graphs with order $n=1,000$ vs.\ the edge density $s$ (on a log scale).  Three conclusions are evident from the plots. First, the bounds on SSR (\prpref{lcssrer}, top left plot) are not as tight as for the unit cost case, but both bounds are tight for separate ranges of values of $s$, and the approximation for SSC (\prpref{lcsscer}, top right plot) and SSS (\prpref{lcssser}, bottom left plot) are ``accurate''. 
Second, the average linear cost ordering on ER graphs is $c_l^{\rm SSR} \geq c_l^{\rm SSC} \geq c_l^{\rm SSS}$ (bottom right plot), as was true for unit cost.  Third, there is a notable difference in linear cost between SSR and SSC/SSS for ``small'' $s$, and a notable difference between SSR/SSC and SSS for ``large'' $s$ (bottom right plot).

\begin{figure}[ht!]
\centering
\includegraphics[width=\columnwidth]{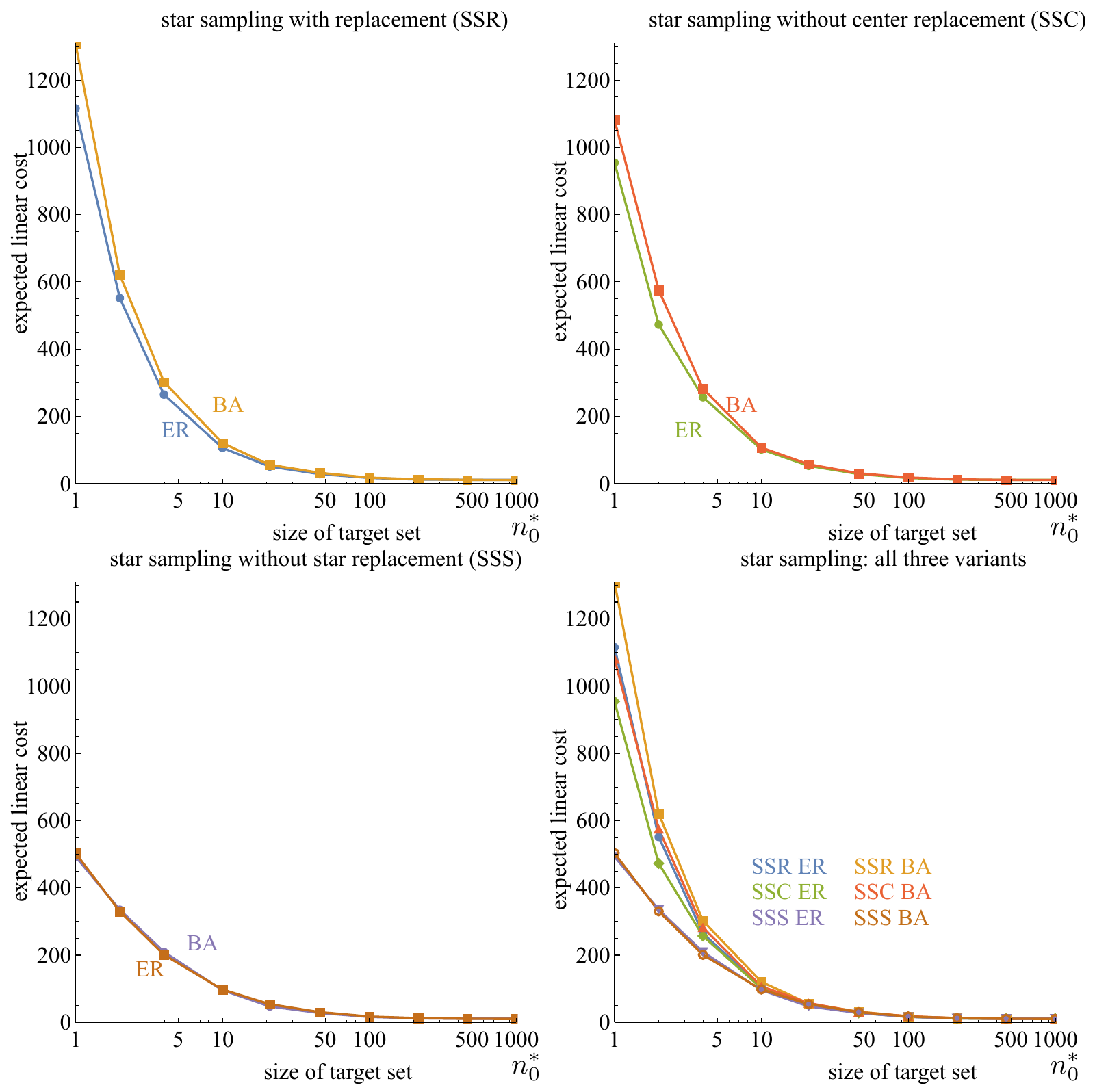}
\caption{Simulation results for expected {\em linear} cost on ER and BA random graphs with order $n=1,000$ and $s = 0.01$ vs.\ target set size $n_0^*$ (on log scale).  {\em Top left:} SSR.  {\em Top right:} SSC.  {\em Bottom left:} SSC.  {\em Bottom right:} all three star sampling types (SSR, SSC, SSS).}
\label{fig:lcerba}
\end{figure}

Recall the discussion regarding \figref{ucerba}; the same simulation parameters given there apply here as well.  The four plots in \figref{lcerba} show simulation results (each point the average over $1,000$ trials) for the expected {\em linear} cost to find $v\in V^*$ under SSR, SSC, and SSS on both ER and BA random graphs with order $n=1,000$ vs.\ the target set size $n_0^*$ (on a log scale).  As with \figref{ucerba}, it is evident from \figref{lcerba} that there is only a small difference in expected linear cost for BA and ER graphs and the BA vs.\ ER linear cost difference decreases in $n_0^*$.  

\section{Results on ``real-world'' graphs}
\label{sec:results}

This section assesses the accuracy when the SSR, SSC, and SSS unit and linear cost estimates, derived for ER graphs in \secref{unit_cost} and \secref{s_cost}, are applied to twelve ``real-world'' graphs. These graphs come from a variety of sources: {\tt power-network} is from \cite{WatStro1998}; {\tt condMat}, {\tt gnutella08}, {\tt gnutella04}, {\tt web-google}, {\tt astroPh}, {\tt epinions}, {\tt fb-combined}, {\tt oregon1}, and {\tt brightkite} are from the SNAP repository \cite{Snap2014}; while the {\tt web-edu} and {\tt tech-routers} are from the network data repository \cite{RosAhm2015}. 

\begin{figure}[ht!]
\begin{center}
\includegraphics[width=0.49\columnwidth]{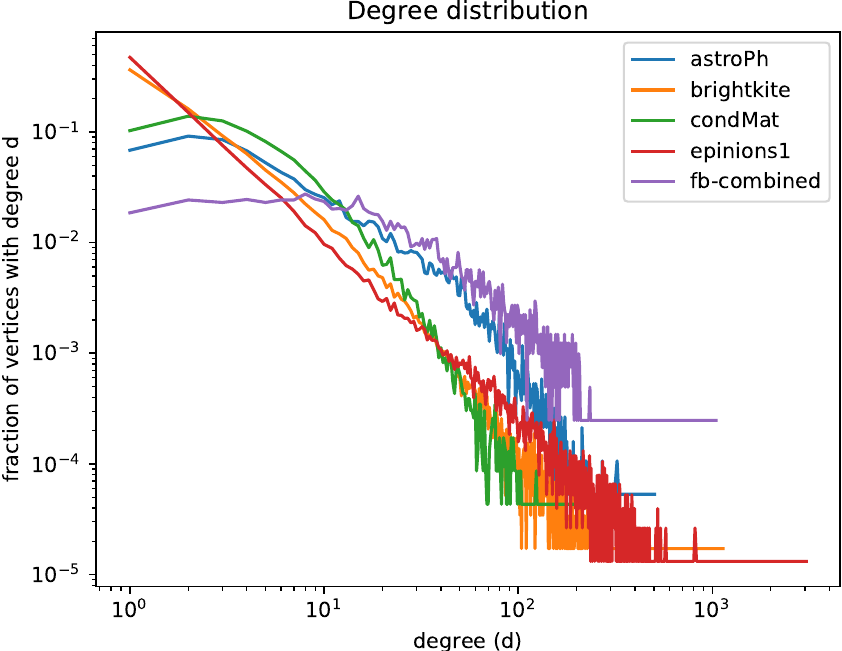}%
\includegraphics[width=0.49\columnwidth]{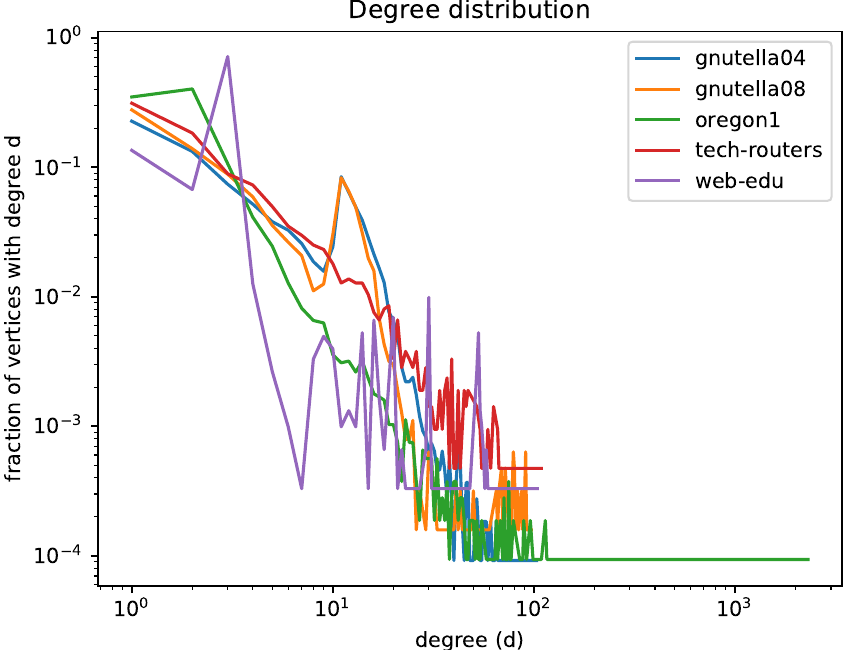}
\caption{Degree distribution for ten of the ``real-world'' graphs.}
\end{center}
\label{fig:8}
\end{figure}

\tabref{graph_stats} gives statistics for these graphs, including order $n \equiv |V|$, size $m \equiv |E|$, edge density $s \equiv m/\binom{n}{2}$, assortativity $\alpha$ (the correlation of the degrees of a randomly selected edge), and maximum degree $d_{\rm max} \equiv \max D$.  The degree distribution (on a log-log scale) for ten of the graphs is shown in \figref{8}; the approximately straight line for some of the graphs (e.g., \texttt{epinions1} and \texttt{brightkite}) suggests some of the graphs have a power-law degree distribution (a key characteristic of the Barab\'{a}si-Albert (BA) graphs studied in \figref{ucerba} and \figref{lcerba}).

\tabref{unit_MD_RND} and \tabref{lin_MD_RND} present empirical and estimated costs, unit and linear respectively, on these graphs under SSR, SSC, and SSS, assuming $n_0^* = 4$, and assuming $\nsf_0^{e,*}$ is known.  Simulation results are presented as a $95\%$ confidence interval after $1,000$ independent trials.  The percent relative error between the simulation mean cost and the estimated expected cost are given for all three sampling variants.  The relative error column entries are bolded when the estimate does not lie in the $95\%$ confidence interval from simulation.  

Several comments bear mention.  First, the SSR and SSC approximations are remarkably accurate, all within $5\%$ relative error.  Second, the SSS approximations, while less accurate than SSR and SSC, are still remarkably accurate given $i)$ these graphs are quite different from one another and quite different from an ER graph, and $ii)$ the approximate costs require so few summary statistics about the graph, i.e., $(n, n_0^*, s)$.  Both unit and linear SSS relative errors are all under $30\%$, several having much smaller relative error, aside from the $~60\%$ error for \texttt{epinions1} (unit) and \texttt{oregon1} (linear). 

\begin{table}
\centering
\begin{tabular}{lrrrrr} 
Graph				& $n$		& $m$		& $s$		& $\alpha$	& $d_{\rm max}$	\\ \hline \hline
\texttt{astroPh}		& 18,772 	& 198,110	& 0.00112	& 0.21		& 504 \\
\texttt{brightkite}		& 58,228	& 214,078	& 0.00013	& 0.11		& 1134 \\
\texttt{condMat}		& 23,133	& 93,497	& 0.00035	& 0.14		& 281 \\
\texttt{epinions} 		& 75,879	& 405,740	& 0.00014	& -0.04		& 3044 \\
\texttt{fb-combined}	& 4,039 	& 88,234	& 0.01082 	& 0.06		& 1045 \\
\texttt{gnutella04}		& 10,876 	& 39,994	& 0.00068 	& -0.01		& 103 \\
\texttt{gnutella08}		& 6,301		& 20,777	& 0.00105	& 0.04		& 97 \\
\texttt{oregon1}		& 10,670	& 22,002	& 0.00039	& -0.19		& 2312 \\
\texttt{power-network}	& 4,941		& 6,594		& 0.00054	& 0.00		& 19 \\
\texttt{tech-routers}	& 2,113 	& 6,632		& 0.00297	& 0.02		& 109 \\
\texttt{web-edu}		& 3,031		& 6,474		& 0.00141	& -0.17		& 104 \\
\texttt{web-google}		& 1,299 	& 2,773		& 0.00329	& -0.05		& 59 \\ \hline  
\end{tabular}
\caption{Statistics of the ``real-world'' graphs.}
\label{tab:graph_stats}
\end{table}

\begin{table*}
\centering
\begin{tabular}{l|r|c|r|r|c|r|r|c|r} \hline
			&SSR-E	&SSR-C	 		&SSR-R	&SSC-E	&SSC-C			&SSC-R	&SSS-E			&SSS-C			&SSS-R\\ \hline
\texttt{astroPh} 	&399.4		&348.0, 410.9	&5.3	&391.1 	&361.5, 424.0	&0.4	&381.1		&369.4, 430.4	&4.7\\
\texttt{britekite}	&1164.6	&1049.2, 1231.4	&2.1	&1141.7 &1070.2, 1260.4	&2.0	&\textbf{1134.6}&1279.1, 1522.5	&19.0\\ 
\texttt{contMat} 	&545.2	&547.4, 644.7	&5.3	&550.8	&546.0, 643.5	&7.4	&545.0 			&508.6, 591.0	&0.9\\
\texttt{epinions1} 	&465.5 	&426.1, 497.8	&1.4	&465.5 	&418.2, 488.1	&2.7	&\textbf{465.1} &1020.5, 1298.3	&60.0\\ 
\texttt{fb-combined} &28.0 	&26.0, 30.4		&0.7	&27.9 	&25.8, 30.2		&0.5	&27.7 			&28.9, 34.3		&12.3\\
\texttt{gnutella04} &302.1	&268.7, 318.5	&2.9	&294.0 	&279.1, 326.2	&2.9	&290.4 			&288.9, 245.7	&8.5\\
\texttt{gnutella08} &525.1 	&479.5, 562.7	&0.8	&484.8 	&461.0, 537.5	&2.9	&\textbf{444.9} &459.0, 537.0	&10.6\\ 
\texttt{oregon1} 	&304.9 	&273.4, 320.5	&2.7	&296.4 	&261.1, 308.0	&4.2	&\textbf{294.4} &343.0, 403.0	&21.1\\ 
\texttt{power-network} &308.8 &270.1, 319.1	&4.8	&290.7 	&266.9, 311.8	&0.5	&285.2 			&248.4, 287.4	&6.4\\
\texttt{tech-routers} &192.0 &176.6, 206.6	&0.2	&176.2 	&168.8, 196.5	&3.6	&\textbf{160.4} &226.9, 261.5	&34.3\\ 
\texttt{web-edu} 	&216.5 	&214.1, 251.5	&7.0	&202.1 	&192.7, 223.4	&2.8	&\textbf{194.0} &226.0, 262.5	&20.6\\ 
\texttt{web-google} 	&86.6	&76.1, 90.2		&4.1	&81.3 	&77.2, 89.7		&2.6	&78.3			&69.6, 80.0		&4.7\\ \hline
\end{tabular}
\caption{Unit cost, $1000$ trials $n_0^*=4$: SSR-E $\to$ SSR Estimate, SSR-C $\to$ SSR $95\%$ Confidence Interval, SSR-R $\to$ SSR Absolute Relative Error $(\%)$, Etc. Bold indicates the estimate is outside the confidence interval.}
\label{tab:unit_MD_RND}
\end{table*}

\begin{table*}
\centering
\begin{tabular}{l|r|c|r|r|c|r|r|c|r} \hline
			&SSR-E		&SSR-C	 		&SSR-R	&SSC-E	&SSC-C			&SSC-R	&SSS-E				&SSS-C			&SSS-R\\ \hline
\texttt{astroPh} 	&8829.6		&7678.2, 9066.4	&5.5	&8477.6 &7852.1, 9174.3	&0.4	&\textbf{5959.2}	&4271.7, 4749.2	&32.1\\
\texttt{britekite} 	&9727.7	 	&8753.0, 10275.4&2.2	&9375.6 &8752.2, 10279.0&1.5	&\textbf{8309.0} 	&6155.9, 7005.0	&26.3\\
\texttt{contMat} 	&5125.0		&4974.1, 5858.9	&5.4	&4899.7	&4849.9, 5693.8	&7.1	&\textbf{4174.0} 	&3435.9, 3901.4 &13.5\\ 
\texttt{epinions1} 	&5477.5 	&4981.9, 5782.3	&1.8	&5413.6 &4885.4, 5669.5	&2.6	&5069.6 			&4944.2, 5765.5	&4.8\\
\texttt{fb-combined} &1253.5 	&1165.7, 1361.4	&0.8	&1236.8 &1147.9, 1334.7	&0.4	&\textbf{928.8} 	&816.5, 915.2	&7.3\\
\texttt{gnutella04} 	&2524.0		&2245.2, 2660.5	&2.9	&2399.1 &2278.6, 2652.2	&2.7	&\textbf{2034.3} 	&1738.3, 1980.4	&9.4\\
\texttt{gnutella08} 	&3987.9 	&3648.3, 4280.2	&0.6	&3453.3 &3293.7, 3801.3	&2.7	&\textbf{2388.9} 	&1805.8, 2006.7	&25.3\\
\texttt{oregon1} 	&1562.1 	&1388.9, 1632.3	&3.4	&1485.8 &1320.1, 1545.9	&3.7	&\textbf{1354.9} 	&793.9, 902.8	&59.7\\
\texttt{power-network} &1133.1 	&990.2, 1169.6	&4.9	&1023.5 &941.8, 1094.7	&0.5	&912.5 				&812.1, 926.4	&5.0\\
\texttt{tech-routers} &1397.9 	&1287.7, 1503.6	&0.2	&1196.8 &1154.1, 1328.5	&3.6	&\textbf{819.7} 	&717.0, 792.0	&8.6\\
\texttt{web-edu} 	&1141.4 	&1137.1, 1335.5	&7.7	&1011.6 &972.0, 1117.5	&3.2	&815.6 				&725.3, 820.5	&5.5\\
\texttt{web-google} 	&456.3		&401.1, 475.1	&4.2	&407.7  &388.0, 448.6	&2.5	&\textbf{330.8} 	&265.6, 296.3 	&17.7\\ \hline
\end{tabular}
\caption{Linear cost, $1000$ trials, $n_0^*=4$: SSR-E $\to$ SSR Estimate, SSR-C $\to$ SSR $95\%$ Confidence Interval, SSR-R $\to$ SSR Absolute Relative Error $(\%)$, Etc. Bold indicates the estimate is outside the confidence interval.}
\label{tab:lin_MD_RND}
\end{table*}

\begin{figure}[ht!]
\centering
\includegraphics[width=\columnwidth]{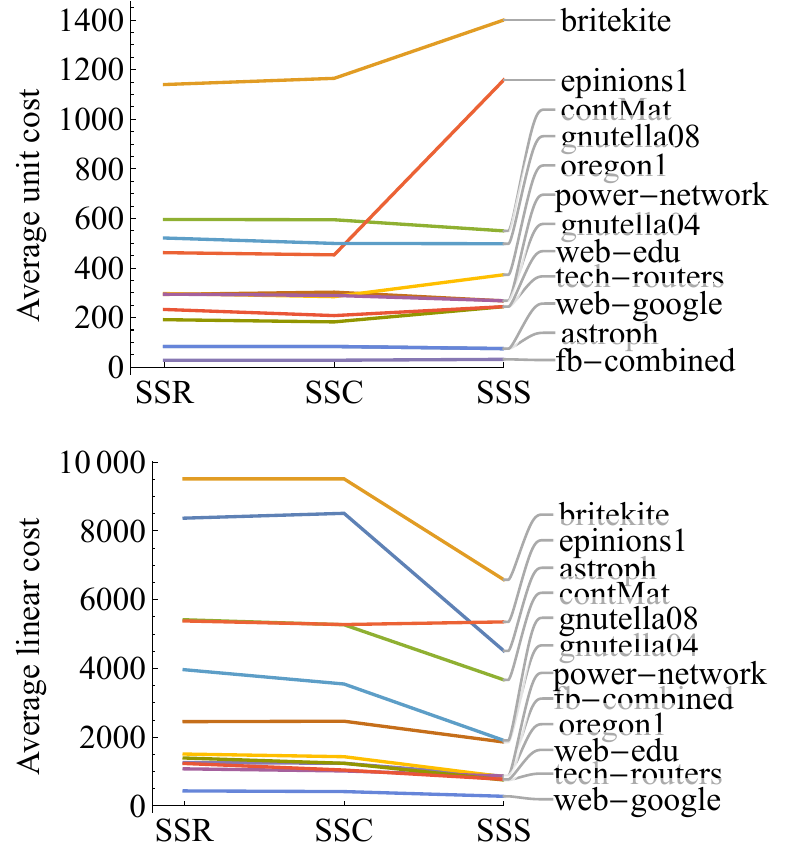}
\caption{Simulation results for the twelve ``real-world'' graphs.}
\label{fig:9}
\end{figure}

\figref{9} shows the unit (top) and linear (bottom) costs for the twelve ``real-world'' graphs vs.\ the sampling variant, i.e., SSR, SSC, or SSS.  The cost for most graphs is relatively constant across the three variants, with several notable exceptions.  Namely, the unit costs for \texttt{epinions1} and \texttt{britekite} are notably higher for SSS than for SSR/SSC, and the linear costs for \texttt{britekite} and \texttt{astroph} are notably lower for SSS than for SSR/SSC.   

\section{Conclusion and discussion}
\label{sec:conclusion}

Star sampling (SS) is a natural means by which to randomly search for an element of a target set in a large / dynamic graph that is not available for deterministic search.  Assessing the performance of the three SS variants (SSR, SSC, and SSS) is an important question to answer in order to assess whether the complexity incurred in avoiding sample replacement will yield a notable reduction in cost. 

The analytical and simulation results in \secref{unit_cost} and \secref{s_cost} suggest the answer to this question is, in general, difficult to predict.  First, consider the unit cost results in \figref{ucnr} which shows, for ER graphs, there is little to no cost difference between SSC and SSS, while there is a substantial cost difference between SSR and SSC for low-density (sparse) graphs.  Second, consider the linear cost results in \figref{lcnr} which shows, again for ER graphs, there is again a cost difference between SSR and SSC for sparse graphs (but not for dense graphs), and moreover there is a cost difference between SSC and SSS for dense graphs (but not for sparse graphs).  With these observations in mind, consider the ``real-world'' graph results of \figref{9} in \secref{results} and note the edge densities for these graphs in \tabref{graph_stats} are in the ``sparse regime'' of \figref{ucnr} and \figref{lcnr}. \figref{9} shows: $i)$ almost no difference in unit or linear cost between SSR and SSC (whereas sparse ER graphs showed a notable SSR to SSC unit and linear difference) and $ii)$ some graphs have {\em higher} unit and {\em lower} linear cost for SSS vs.\ SSC (whereas sparse ER graphs showed a negligible SSC to SSS unit and linear cost difference).  

The substantial {\em increase} in unit cost from SSC to SSS for the \texttt{britekite} and \texttt{epinions1} graphs in \figref{9} merits comment, as all numerical results for ER graphs in \secref{unit_cost} and \secref{s_cost} consistently show SSS to have {\em lower} cost than SSC.  Intuition suggests that SSS will have lower expected cost than SSC as removing an entire star removes more non-target vertices than does removing a non-target star center.  While this intuition may hold in some cases, it does not hold in general, as the unit costs for \texttt{britekite} and \texttt{epinions1} graphs demonstrate.  One possible explanation for the increase is that removing ``large'' non-target stars may end up removing more of the {\em neighbors} of the target set than removing a non-target star center, and the reduction in the number of target set neighbors will decrease the hitting probability for the subsequent sample.

The results in \tabref{unit_MD_RND} and \tabref{lin_MD_RND} show that the unit and linear cost estimators under both SSR and SSC are quite accurate, while the unit and linear cost estimators under SSS are subject to sometimes substantial error.  

Given the above three paragraphs, it is evidently difficult to provide simple, accurate, and widely-applicable conclusions concerning the theoretical and empirical results.  As mentioned in the introduction, there is cause for considering both cost models, as star samples of graphs where each vertex has (does not have) access to the degree of its neighbors will have unit (linear) cost, respectively.  Likewise, the ``relative cost'' of the three sampling paradigms will depend upon the specifics of how the graph is accessed: while SSR is always ``simpler'' than SSC, and SSC is always ``simpler'' than SSS, the added complexity required to remove star centers and or stars from the list of eligible vertices for sampling will depend upon the context.  Having chosen a cost model and a star sampling paradigm, the accuracy of the cost estimator is variable.  

Despite these complexities and nuances, it nonetheless remains the case that the results provide reasonably accurate approximations on expected costs for two cost models and for three sampling paradigms in terms of widely available graph statistics.  Future work may refine the analysis technique to yield more accurate and/or more widely applicable results.  

\section*{Acknowledgment}

The authors wish to express their thanks to the Anonymous Reviewers for their substantive and helpful suggestions.

\bibliographystyle{IEEEtran}
\bibliography{IEEEabrv,TCSS2019sub}

\appendices

\section{Proof of \facref{ecnra}}
\label{app:ecnra}

\begin{IEEEproof}
Define:
$i)$ $(n)_k \equiv (n-0)(n-1)\cdots(n-(k-1))$ as the falling factorial (recall $\binom{n}{k} = (n)_k/k!$);
$ii)$ $p_t \equiv n^*/(n-t+1)$ for $t \in [n-n^*+1]$ as the probability of success on trial $t$, conditioned on failure in the first $t-1$ trials (observe $p_1 = n^*/n$ and $p_{n-n^*+1} = 1$);  
$iii)$ $q_t \equiv p_t \prod_{c=1}^{t-1} (1-p_c)$ as the unconditioned probability of a first success on trial $t$;
$iv)$ $\xsf$ as the RV for the number of required trials, where $\xsf$ has distribution $(q_t, t \in [n-n^*+1])$.  It is straightforward but tedious to show that $(q_t, t \in [n-n^*+1])$ is correctly normalized.  Substituting into the definition $\Ebb[\xsf] \equiv \sum_{t=1}^{n+1-n^*} t q_t$ yields
\begin{equation}
\Ebb[\xsf] = \frac{n^*}{n} + n^* \sum_{t=2}^{n+1-n^*} \frac{t}{n+1-t} \prod_{s=1}^{t-1} \frac{n+1-n^*-s}{n+1-s}.
\end{equation}
Using the falling factorial definition yields:
\begin{equation}
\Ebb[\xsf] = \frac{n^*}{n} + n^* \sum_{t=2}^{n+1-n^*} \frac{\binom{n-n^*}{t-1}}{\binom{n}{t}}\nonumber \\
\end{equation}
Using the change of variable $s=n+1-n^*-t$ yields 
\begin{equation}
\Ebb[\xsf] = \frac{n^*}{n} + \frac{1}{\binom{n}{n^*}} \sum_{s=0}^{n-1-n^*} (n+1-n^*-s) \binom{n^*-1+s}{n^*-1}. 
\end{equation}
It is straightforward but tedious to verify that 
\begin{equation}
\sum_{s=0}^{n-1-n^*} (n+1-n^*-s) \binom{n^*-1+s}{n^*-1} = \frac{(n+n^*+1)(n-1)!}{(n-n^*-1)!(n^*+1)!}.
\end{equation}
Substitution and simple algebra yields the result.
\end{IEEEproof}

\section{Proof of \lemref{tbd1}}
\label{app:tbd1}

\lemref{tbd1} holds by \lemref{erssc} and \lemref{sdge}, both proved below.

\begin{lemma}
\label{lem:erssc}
Take $t' \in [n]$ samples from an ER random graph $\Gsf$ with parameters $(n,s)$ using SSC, yielding a sequence of random graphs $(\Gsf_t, t \in [t'])$.  Then $\Gsf_t = (V_t,\Esf_t)$ is an ER random graph with parameters $(n-t,s)$ for each $t \in [t']$.
\end{lemma}

\begin{IEEEproof}
Let $\Gsf_0 = \Gsf$ denote the initial ER random graph, with $\Gsf_0 = (V_0,\Esf_0)$, where $V_0 = [n]$ and $\Esf_0$ may be considered as a sequence of $\binom{n}{2}$ IID Bernoulli RVs, say $\Esf_0 = (\xsf_e, e \in [\binom{n}{2}])$, with $\xsf_e \sim \mathrm{Ber}(s)$ indicating the random inclusion or exclusion of an edge at the ``site'' at (unordered) vertex pair $e$.

Define the sequence of random vertex and set triples $((\vsf_t,\Vsf_t,\Esf_t), t \in [t'])$, with $\vsf_t \sim \mathrm{uni}(\Vsf_{t-1})$ the randomly selected star center in sample $t$, $\Vsf_t = \Vsf_{t-1} \setminus \{\vsf_t\}$ the set of vertices that remain after removal of star center $\vsf_t$, and $\Esf_t = \Esf_{t-1} \setminus \Fsf_t$, for $\Fsf_t \equiv \Gamma_{\Gsf_{t-1}}(\vsf_t)$, the edges that remain after removal of the edge neighborhood of $\vsf_t$ from $\Gsf_{t-1}$.

For $t \in [t']$: $i)$ $\Vsf_t = V_0 \setminus \Vsf_t^c$, for $\Vsf_t^c \equiv \vsf_1 \cup \cdots \cup \vsf_t$ the set of vertices removed in the first $t$ samples, and $ii)$ $\Esf_t = \Esf_0 \setminus \Esf_t^c$, for $\Esf_t^c \equiv \Fsf_1 \cup \cdots \cup \Fsf_t$ the set of edges removed in the first $t$ samples.  Then $|\Vsf_t| = n-t$, and, as $\Esf_0$ is IID $\mathrm{Ber}(s)$, the RVs in $\Esf_t$ are IID $\mathrm{Ber}(s)$, so that $\Gsf_t$ is an ER random graph (with random vertex labels) with parameters $(n-t,s)$.
\end{IEEEproof}

\begin{lemma}
\label{lem:sdge}
Take $t' \in [n]$ samples from an ER random graph $\Gsf$ with parameters $(n,s)$ using SSS, yielding a sequence of random graphs $(\Gsf_t, t \in [t'])$.  Then $\Gsf_t = (\Vsf_t,\Esf_t)$ is an ER random graph with random order parameter $\nsf_t$ and edge probability $s$, for each $t \in [t']$.  The random order parameter, $\nsf_t \equiv |\Vsf_t|$, obeys the recursion $\nsf_t = \nsf_{t-1} - \dsf^e_t$, for $\dsf^e_t \sim 1+\mathrm{bin}(\nsf_{t-1}-1,s)$, for $t \in [t']$.
\end{lemma}

\begin{IEEEproof}
The first two paragraphs in the proof of \lemref{erssc} hold here as well; we adopt the same notation below. Define the sequence of random vertex and set triples $((\vsf_t,\Vsf_t,\Esf_t), t \in [t'])$, with $\vsf_t \sim \mathrm{uni}(\Vsf_{t-1})$ the randomly selected star center in sample $t$, $\Vsf_t = \Vsf_{t-1} \setminus N_{\Gsf_{t-1}}^e(\vsf_t)$ the set of vertices that remain after removal of {\em star} $N_{\Gsf_{t-1}}^e(\vsf_t)$, and $\Esf_t = \Esf_{t-1} \setminus \Gamma_{\Gsf_{t-1}}^e(\vsf_t)$, the edges that remain after removal of the {\em extended} edge neighborhood of $\vsf_t$ from $\Gsf_{t-1}$.  The sequence of RV pairs $((\nsf_t,\dsf^e_t), t \in [t'])$, where $\nsf_t = |\Vsf_t|$ is the random order of $\Gsf_t$ and $\dsf^e_t = |N_{\Gsf_{t-1}}^e(\vsf_t)|$ is the random order of the extended neighborhood of $\vsf_t$, obeys (by construction) the recursion $\nsf_t = \nsf_{t-1} - \dsf^e_t$, with $\nsf_0 = n$.  Proof by induction in $t$ will show $i)$ $\Gsf_t$ is ER, and $ii)$ $\dsf^e_t \sim 1+\mathrm{bin}(\nsf_{t-1}-1,s)$, for $t \in [t']$. 

{\em Base case.} It must be shown $i)$ $\Gsf_1 = (\Vsf_1,\Esf_1)$ is an ER random graph with parameters $\nsf_1 = n - \dsf^e_1$ and $s$, and $ii)$ $\dsf^e_1 \sim 1 + \mathrm{bin}(n-1,s)$.  Recall $\Vsf_1 = V_0 \setminus N^e_{G_0}(\vsf_1)$ and $\Esf_1 = \Esf_0 \setminus \Gamma_{G_0}^e(\vsf_1)$.  First, the edges in $\Esf_1$ are a (randomly selected) subset of the IID $\mathrm{Ber}(s)$ RVs in $\Esf_0$, and as such the edges in $\Esf_1$ are also IID $\mathrm{Ber}(s)$ RVs; this shows $i)$.  Second, $\dsf^e_1 = |N_{G_0}^e(\vsf_1)|$ counts both the star center $\usf_1$ and its neighbors $N_{G_0}(\vsf_1)$, where $d_{\vsf_1} \equiv |N_{G_0}(\vsf_1)|$ has distribution $\mathrm{Ber}(n-1,s)$ by construction; this shows $ii)$.  

{\em Induction hypothesis.}  Fix $t \in [t']$ and assume the induction hypothesis for $t-1$, i.e., assume $i)$ $\Gsf_{t-1}$ is an ER random graph with parameters $\nsf_{t-1}$ and $s$, and $ii)$ $\dsf^e_{t-1} \sim 1+\mathrm{bin}(\nsf_{t-2}-1,s)$.  It follows that $\nsf_{t-1} = \nsf_{t-2} - \dsf^e_{t-1}$. It must shown that this implies $i)$ the random graph $\Gsf_t$ obtained by removing a random star from $\Gsf_{t-1}$ is an ER graph with parameters $\nsf_t = \nsf_{t-1} - \dsf^e_t$ and $s$, and $ii)$ $\dsf^e_t \sim 1+\mathrm{bin}(\nsf_{t-1},s)$.  First, the edges in $\Esf_t$ are a (randomly selected) subset of the IID $\mathrm{Ber}(s)$ RVs in $\Esf_{t-1}$, and as such the edges in $\Esf_t$ are also IID $\mathrm{Ber}(s)$ RVs; this shows $i)$.  Second, $\dsf^e_t = |N_{\Gsf_{t-1}}^e(\vsf_t)|$ counts both the star center $\vsf_t$ and its neighbors $N_{\Gsf_{t-1}}(\vsf_t)$, where $d_{\vsf_t} \equiv |N_{\Gsf_{t-1}}^e(\vsf_t)|$ has distribution $\mathrm{Ber}(\nsf_{t-1}-1,s)$ by construction; this shows $ii)$.
\end{IEEEproof}

\section{Proof of \lemref{gosm}}
\label{app:gosm}

\begin{IEEEproof}
A star sample has two ``types'' of vertices (the star center $\vsf$ and its $\dsf$ neighbors) and three ``types'' of edges, namely, $i)$ ``neighbor'' edges connecting $\vsf$ with $N_{\Gsf}(\vsf)$, $ii)$ ``internal'' edges with both vertices in $N_{\Gsf}(\vsf)$, and $iii)$ ``external'' edges with one vertex in $N_{\Gsf}(\vsf)$ and the other in $V \setminus N_{\Gsf}^e(\vsf)$.  There are $\dsf$ neighbor edges by assumption, $\binom{\dsf}{2}$ potential internal edges, and $\dsf(n - \dsf - 1)$ potential external edges, where (due to the ER random graph properties) all potential edges are present or absent independently with probability $s$.  

As $\dsf \sim \mathrm{bin}(n-1,s)$, we have $\Ebb[\dsf] = (n-1)s$ and $\Ebb[\dsf^2] = (n-1)s + 2 \binom{n-1}{2} s^2$, and with these \eqnref{expgsfnumedgesss} is obtained by conditional expectation and simple algebra:
\begin{eqnarray}
\Ebb[\gsf]
&=& \Ebb[\Ebb[\gsf|\dsf]] \nonumber \\
&=& \Ebb\left[\dsf + \left(\binom{\dsf}{2} + \dsf(n-\dsf-1)\right)s \right] \nonumber \\
&=& \Ebb\left[\left(1 + \left(n-\frac{3}{2}\right)s \right)\dsf - \frac{1}{2}s \dsf^2 \right] \nonumber \\
&=& \left(1 + \left(n-\frac{3}{2}\right)s \right)\Ebb[\dsf] - \frac{1}{2}s \Ebb[\dsf^2] \end{eqnarray}
The limit of $\Ebb[\gsf]/(\binom{n}{2}s)$ as $n \uparrow \infty$ follows from \eqnref{expgsfnumedgesss}.
\end{IEEEproof}

\section{Proof of \lemref{watchdraw}}
\label{app:watchdraw}

Let $(\Gsf_t, t \in \Zbb^+)$ be a sequence of ER random graphs induced by SSS, with $\Gsf_t = (\Vsf_t,\Esf_t)$ the graph following the removal of the random star selected in sample $t$, starting from an initial ER random graph $\Gsf_0 = (V_0,\Esf_0)$ with parameters $(n,s)$.  Set $(1-s) \equiv 1-s$.  

Fix ``watch'' $W_0 \subseteq V_0$ and ``draw'' $D_0 \subseteq V_0$ subsets, and define an ``immune'' subset $Z_0 \subseteq W_0$ comprised of vertices in $W_0$ with no neighbors in $D_0$.  Define the random set-valued sequences $(\Wsf_t, t \in \Zbb^+)$ and $(\Dsf_t, t \in \Zbb^+)$, with $\Wsf_t \equiv W_0 \cap \Vsf_t$ and $\Dsf_t \equiv D_0 \cap \Vsf_t$ the watch and draw sets surviving the first $t$ samples. Consider two cases for star centers: $i)$ they are {\em never} drawn from the watch set, i.e., $D_0 \cap W_0 = \emptyset$, and $ii)$ they are {\em always} drawn from the watch set, i.e., $D_0 \subseteq W_0$.  

Let $(\vsf_t, t \in \Nbb)$, with $\vsf_t \sim \mathrm{uni}(\Vsf_{t-1})$, denote the sequence of random star centers associated with each random sample, and define the event sequence $(\Cmc_t, t \in \Nbb)$ that star center $t$ is drawn from the surviving draw set, i.e., $\Cmc_t \equiv \{\vsf_t \in \Dsf_{t-1}\}$.  Let $\bar{\Cmc}_t \equiv \bigcap_{t' \in [t]} \Cmc_{t'}$ be the event that the first $t$ star centers are each drawn from the surviving draw set.  

Define the random sequence $(\nsf_t, t \in \Zbb^+)$, with $\nsf_t \equiv |\Wsf_t|$ the watch set order, and define the sequences of conditional means $(\mu_{W|D,t}, t \in \Zbb^+)$ and conditional variances $(\sigma^2_{W|D,t}, t \in \Zbb^+)$, with $\mu_{W|D,t} \equiv \Ebb[\nsf_t|\bar{\Cmc}_t]$ and $\sigma^2_{W,t} \equiv \mathrm{Var}(\nsf_t|\bar{\Cmc}_t)$ associated with $\nsf_t$ and $\bar{\Cmc}_t$.  Let $n_{Z,0} = |Z_0|$.

\begin{lemma}
\label{lem:watchdraw}
The mean and variance in the size of the watch set $\Wsf_t$, conditioned on drawing from the draw set $\Dsf_t$ and removing no vertices from the immune set $Z_0$, after $t$ samples, where $n_{W,0}=|W_0|$ and $n_{Z,0}=|Z_0|$, are: 
$i)$ if $D_0 \cap W_0 = \emptyset$,
\begin{eqnarray}
\mu_{W|D,t}^{i)} &=& (n_{W,0}-n_{Z,0}) (1-s)^t + n_{Z,0} \label{eq:muWDtc1} \\
\sigma^{2,i)}_{W|D,t} &=& (n_{W,0}-n_{Z,0}) (1-s)^t (1 - (1-s)^t) \label{eq:sigma2WDtc1}
\end{eqnarray}
or, $ii)$ if $D_0 \subseteq W_0$ and $Z_0 = \emptyset$, 
\begin{eqnarray}
\mu_{W|D,t}^{ii)} &\!\!\!=\!\!\!& \mu_{W|D,t}^{i)} - \frac{(1-s)(1-(1-s)^t)}{s} \label{eq:muWDtc2} \\
\sigma^{2,ii)}_{W|D,t} &\!\!\!=\!\!\!& \sigma^{2,i)}_{W|D,t} + \frac{(1-s)^{t+1}\left(1 + (1-s)(1-(1-s)^t) \right)}{s(2-s)}. \label{eq:sigma2WDtc2} 
\end{eqnarray}
\end{lemma}

\begin{IEEEproof}
Let $\rsf_t$ be the RV denoting the number of vertices in the watch set {\em removed} by sample $t$.  The random recurrence induced by the sampling is $\nsf_t = \nsf_{t-1} - \rsf_t$.  First:  
\begin{eqnarray}
\mu_{W|D,t}
& \equiv & \Ebb[\nsf_t | \bar{\Cmc}_t] = \Ebb[\Ebb[\nsf_t|\nsf_{t-1}] | \bar{\Cmc}_t] \\
&=& \Ebb[\Ebb[\nsf_{t-1} - \rsf_t|\nsf_{t-1}] | \bar{\Cmc}_t] = \Ebb[\nsf_{t-1} - \Ebb[\rsf_t|\nsf_{t-1}] | \bar{\Cmc}_t]. \nonumber 
\end{eqnarray}
Second: $\sigma^2_{W|D,t}$
\begin{eqnarray}
\label{eq:ypdo}
& \equiv & \mathrm{Var}(\nsf_t | \bar{\Cmc}_t) \nonumber \\
& = & \mathrm{Var}(\Ebb[\nsf_t|\nsf_{t-1}] | \bar{\Cmc}_t)+ \Ebb[\mathrm{var}(\nsf_t|\nsf_{t-1})|\bar{\Cmc}_t] \nonumber \\
& = &\mathrm{Var}(\Ebb[\nsf_{t-1} - \rsf_t|\nsf_{t-1}] | \bar{\Cmc}_t) + \Ebb[\mathrm{Var}(\nsf_{t-1} - \rsf_t|\nsf_{t-1})|\bar{\Cmc}_t] \nonumber \\
& = & \mathrm{Var}(\nsf_{t-1} - \Ebb[\rsf_t|\nsf_{t-1}] | \bar{\Cmc}_t) + \Ebb[\mathrm{Var}(\rsf_t|\nsf_{t-1})|\bar{\Cmc}_t] 
\end{eqnarray}

{\em Case $i)$} ($D_0 \cap W_0 = \emptyset$).  As the star center is {\em never} drawn from the watch set $\rsf_t | (\nsf_{t-1},\Cmc_t) \sim \mathrm{bin}(\nsf_{t-1} - n_{Z,0},s)$, and therefore $i)$ $\Ebb[\rsf_t | \nsf_{t-1}, \Cmc_t] = (\nsf_{t-1} - n_{Z,0})s$ and $ii)$ $\mathrm{Var}(\rsf_t|\nsf_{t-1}, \Cmc_t) = (\nsf_{t-1} -n_{Z,0}) s (1-s)$.  First: 
\begin{eqnarray}
\mu_{W|D,t} - n_{Z,0} &=& \Ebb[(\nsf_{t-1} -n_{Z,0}) - (\nsf_{t-1} -n_{Z,0})s | \bar{\Cmc}_t] \nonumber\\
&=& (1-s)\Ebb[(\nsf_{t-1} -n_{Z,0}) | \bar{\Cmc}_t] \nonumber\\
&=& (1-s)\Ebb[(\nsf_{t-1} -n_{Z,0}) | \bar{\Cmc}_{t-1}] \nonumber\\
&=& (1-s)(\mu_{W|D,t-1} -n_{Z,0}). 
\end{eqnarray}
The recurrence $\mu_{W|D,t} - n_{Z,0} = (1-s) (\mu_{W|D,t-1} - n_{Z,0})$ with initial condition $\mu_{W|D,0} = n_{W,0}$
has solution \eqnref{muWDtc1}.  Next, $\sigma^2_{W|D,t}$
\begin{eqnarray}
&=& \mathrm{Var}((\nsf_{t-1}-n_{Z,0})-(\nsf_{t-1}-n_{Z,0})s | \bar{\Cmc}_t) \nonumber\\
&+& \Ebb[(\nsf_{t-1}-n_{Z,0})s(1-s) | \bar{\Cmc}_{t-1}] \nonumber\\
&=& s^2\mathrm{Var}(\nsf_{t-1}-n_{Z,0}| \bar{\Cmc}_t) + s(1-s)\Ebb[\nsf_{t-1}-n_{Z,0} | \bar{\Cmc}_{t-1}] \nonumber\\
&=& s^2\mathrm{Var}(\nsf_{t-1} | \bar{\Cmc}_{t-1}) + s(1-s)\Ebb[\nsf_{t-1}-n_{Z,0} | \bar{\Cmc}_{t-1}] \nonumber\\
&=& s^2\sigma^2_{W|D,t-1} + s(1-s)\Ebb[\nsf_{t-1}-n_{Z,0} | \bar{\Cmc}_{t-1}] \nonumber\\
&=& s^2\sigma^2_{W|D,t-1} + s(1-s)(\mu_{W|t-1}-n_{Z,0}) \nonumber\\
&=& s^2\sigma^2_{W|D,t-1} + s(1-s)((n_{W,0}-n_{Z,0})(1-s)^{t-1}) \nonumber\\
&=& s^2\sigma^2_{W|D,t-1} + s(1-s)^t(n_{W,0}-n_{Z,0}). 
\end{eqnarray}
The recurrence $\sigma^2_{W|D,t} = (1-s)^2 \sigma^2_{W|D,t-1} + s (1-s)^{t} (n_{W,0} - n_{Z,0})$ with initial condition $\sigma^2_{W|D,0} = 0$ has solution \eqnref{sigma2WDtc1}.

{\em Case $ii)$} ($D_0 \subseteq W_0$, $Z_0 = \emptyset$).  As the star center is {\em always} drawn from the watch set and $n_{Z,0}=0$, $\rsf_t | (\nsf_{t-1},\bar{\Cmc}_t) \sim 1 + \mathrm{bin}(\nsf_{t-1}-1,s)$, and therefore $i)$ $\Ebb[\rsf_t|\nsf_{t-1}] = 1+(\nsf_{t-1}-1) s$ and $ii)$ $\mathrm{Var}(\rsf_t|\nsf_{t-1}) = (\nsf_{t-1}-1) s (1-s)$.  First:
\begin{eqnarray}
\mu_{W|D,t}
&=& \Ebb[\nsf_{t-1} - (1+(\nsf_{t-1}-1) s) | \bar{\Cmc}_t] \nonumber \\
&=& (1-s) \Ebb[\nsf_{t-1} | \bar{\Cmc}_t] - (1-s) \nonumber \\
&=& (1-s) \Ebb[\nsf_{t-1} | \bar{\Cmc}_{t-1}] - (1-s) \nonumber \\
&=& (1-s) (\mu_{W|D,t-1}-1). 
\end{eqnarray}
The recurrence $\mu_{W|D,t} = (1-s) (\mu_{W|D,t-1}-1)$ with initial condition $\mu_{W|D,0} = n_{W,0}$ has solution \eqnref{muWDtc2}. Second: $\sigma^2_{W|D,t}$
\begin{eqnarray}
& = & \mathrm{Var}(\nsf_{t-1} - (1+(\nsf_{t-1}-1) s) | \bar{\Cmc}_t) \\
&+& \Ebb[(\nsf_{t-1}-1) s(1-s)|\bar{\Cmc}_t) \nonumber \\
& = & (1-s)^2 \mathrm{Var}(\nsf_{t-1} | \bar{\Cmc}_t)+ s(1-s)\Ebb[\nsf_{t-1}-1 |\bar{\Cmc}_t) \nonumber \\
& = & (1-s)^2 \mathrm{Var}(\nsf_{t-1} | \bar{\Cmc}_{t-1})+ s(1-s)\Ebb[\nsf_{t-1}-1 |\bar{\Cmc}_{t-1}) \nonumber \\
& = & (1-s)^2 \sigma^2_{W|D,t-1} + s(1-s)(\mu_{W|D,t-1}-1) \nonumber \\
& = & (1-s)^2 \sigma^2_{W|D,t-1} + s \mu_{W|D,t} \nonumber \\
& = & (1-s)^2 \sigma^2_{W|D,t-1} \nonumber \\
& + & (1-s) \left( ((n_{W,0}-1)s + 1) (1-s)^{t-1} - 1 \right). \nonumber
\end{eqnarray}
The recurrence 
\begin{equation}
\sigma^2_{W|D,t} = (1-s)^2 \sigma^2_{W|D,t-1} + ((n_{W,0}-1)s + 1) (1-s)^t - (1-s), \nonumber
\end{equation}
with initial condition $\sigma^2_{W|D,0} = 0$ has solution \eqnref{sigma2WDtc2}.
\end{IEEEproof}

\section{Proof of \thmref{lfeo}}
\label{app:lfeo}

\begin{IEEEproof}
The proof relies upon \lemref{watchdraw} in \appref{watchdraw}.

{\em \eqnref{paucl} through \eqnref{t1tildep}.} 
Apply \lemref{watchdraw} to each of the numerator and denominator in \eqnref{approxpt}. Specifically, $\Ebb[\nsf^{e,*}_t | \bar{\Emc}_t]$ follows from the expectation expression for case $i)$ with watch set $N_{G_0}^e(V^*)$, (disjoint) draw set $V_0 \setminus N_{G_0}^e(V^*)$, conditioned on $\bar{\Emc}_t$ and hence $V^*$ being unsampled, while $\Ebb[\nsf_t | \bar{\Emc}_t]$ follows from the expectation expression for case $ii)$ with watch set $V_0$ and (subset) draw set $V_0 \setminus N_{G_0}^e(V^*)$:
\begin{eqnarray}
\Ebb[\nsf^{e,*}_t | \bar{\Emc}_t] &=& (\Ebb[\nsf^{e,*}_0]-n^*_0) (1-s)^t +n^*_0 \label{eq:ensfesgemc} \\
\Ebb[\nsf_t | \bar{\Emc}_t] &=& n (1-s)^t - \frac{(1-s)}{s} (1 - (1-s)^t). \label{eq:engemc}
\end{eqnarray}
The ratio of \eqnref{ensfesgemc} and \eqnref{engemc} gives \eqnref{paucl}.  The first two derivatives of $\tilde{p}_t^{\rm SSS}$ with respect to $t$ are:
\begin{eqnarray}
\tilde{p}_{t+1}^{\rm SSS'} &=&  \frac{s(1-s)^t(n^*_0ns + \Ebb[\nsf_0^{e,*}](1-s))\log(\frac{1}{(1-s)})}{((1-s)-(1-s)^t(ns+(1-s)))^2} \nonumber \\
\tilde{p}_{t+1}^{\rm SSS''} &=& \frac{s^{t+2}\Ebb[\nsf_0^{e,*}]((1-s)+(1-s)^t(ns+(1-s)))}{((1-s)^t(ns+(1-s))-(1-s))^3} \nonumber \\
& & \times \left(\log{(1-s)}\right)^2.
\end{eqnarray}
Observe both $\tilde{p}_{t+1}^{\rm SSS'} > 0$ and $\tilde{p}_{t+1}^{\rm SSS''} > 0$. This establishes that $\tilde{p}_t^{\rm SSS}$ is convex increasing.  The value $t^{(2)}_{\tilde{p}}$ is found by equating the denominator in $\tilde{p}_t^{\rm SSS}$ to zero and solving for $t$, and $t^{(1)}_{\tilde{p}}$ is found by solving $\tilde{p}_t^{\rm SSS} = 1$ for $t$.

{\em Proof of \eqnref{epsilonntucl}.}
Apply the variance expression in case $ii)$ of \lemref{watchdraw} with watch set $V_0$ and (subset) draw set $V_0 \setminus N_{G_0}^e(V^*)$: $\mathrm{var}(\nsf_t|\bar{\Emc}_t)= (1-s)^t \times$
\begin{equation}
\label{eq:varnsfemct}
\left( n (1 - (1-s)^t) + \frac{(1-s)\left(1 + (1-s)(1-(1-s)^t) \right)}{s(2-s)}\right).
\end{equation}

Next, observe $\mathrm{cov}(\nsf^{e,*}_t,\nsf_t| \bar{\Emc}_t)$ in \eqnref{nfoc} is nonnegative.  To see this, consider a general scenario where each member of a population is given a property value, and two subsets of the population are defined as holding those members with property values in a given target set, with the second target set a subset of the first.  Let the property values be random, and consider the two random variables denoting the cardinalities of the two random population subsets.  These random variables are by construction positively correlated.  This general scenario applies here to the population $\Vsf_t$ with the first subset equal to $\Vsf_t$ and the second subset equal to $N_{\Gsf_t}^e(\Vsf_t^*)$.

By the above argument, \eqnref{nfoc} may be upper bounded as: 
\begin{equation}
\label{eq:nfoc2}
\tilde{\epsilon}_{n,t} \leq \frac{\mathrm{var}(\nsf_t| \bar{\Emc}_t) \Ebb[\nsf^{e,*}_t| \bar{\Emc}_t]}{\Ebb[\nsf_t| \bar{\Emc}_t]^3}.
\end{equation}
Substitution of \eqnref{ensfesgemc}, \eqnref{engemc}, and \eqnref{varnsfemct} into \eqnref{nfoc2} gives \eqnref{epsilonntucl}.

{\em Proof of \eqnref{asymperrorucl}.} This follows by observing the numerator is $O(n n_0^{e,*}) = O(n^2)$ while the denominator is $O(n^3)$.  
\end{IEEEproof}

\section{Proof of \thmref{appuncprobhittarsetSSS}}
\label{app:appuncprobhittarsetSSS}

\begin{IEEEproof}
The proofs of \eqnref{tildeqtsss} and \eqnref{approxexpunitcostundersss} are given in turn.

{\em Proof of \eqnref{tildeqtsss}.} Set $\Cmc_t$ as the event that star $t$ misses the target, $\bar{\Cmc}_t = \Cmc_1 \cap \cdots \cap \Cmc_t$ as the event that the first $t$ stars each miss the target, and $\Cmc_t^c$ as the complement of $\Cmc_t$, i.e., the event that star $t$ hits the target.  Set $p_{t+1} \equiv \Pbb(\Cmc_{t+1}^c|\bar{\Cmc}_t)$ as the {\em conditional} probability that star $t+1$ hits the target {\em given} that the first $t$ stars miss the target.  Note $1-p_{t+1} = \Pbb(\Cmc_{t+1}|\bar{\Cmc}_t)$.  Set $q_{t+1} \equiv \Pbb(\Cmc_{t+1}^c \cap \bar{\Cmc}_t)$ as the {\em unconditioned} probability that star $t+1$ hits the target {\em and} that the first $t$ stars miss the target.  Thus $q_{t+1}$ is the unconditioned probability that the first star to hit the target is star $t+1$.  Set $r_t \equiv \Pbb(\bar{\Cmc}_t)$ as the probability that the first $t$ stars each miss the target.  As
\begin{equation}
q_{t+1} 
= \Pbb(\Cmc_{t+1}^c | \bar{\Cmc}_t) \Pbb(\bar{\Cmc}_t)
= p_{t+1} r_t.
\end{equation}
and
\begin{equation}
r_t 
= \Pbb(\Cmc_1) \Pbb(\Cmc_2 | \Cmc_1) \Pbb(\Cmc_3 | \bar{\Cmc}_2) \cdots \Pbb(\Cmc_t | \bar{\Cmc}_{t-1}) 
= \prod_{s=1}^t (1-p_s),
\end{equation}
it follows that $q_{t+1} = p_{t+1} \prod_{s=1}^t (1-p_s)$.  Given approximation $\tilde{p}_{t+1} \approx p_{t+1}$ of the {\em conditional} distribution, \eqnref{tildeqtsss} is a corresponding approximation of the {\em unconditional} distribution.

{\em Proof of \eqnref{approxexpunitcostundersss}.} Consider a sequence $\hat{p} \equiv (\hat{p}_t, t \in \Nbb)$ with each $\hat{p}_t \in (0,1)$. For $T \in \Nbb$ set $p^{(T)} \equiv (p^{(T)}_t, t \in [T])$, with 
\begin{equation}
p^{(T)}_t \equiv \left\{ \begin{array}{ll}
\hat{p}_t, \; & t \in [T-1] \\
1, \; & t = T 
\end{array} \right.,
\end{equation}
Define the following quantities:
\begin{itemize}
\item $\bar{p}^{(T)}_t \equiv 1 - p^{(T)}_t$;
\item  $\pi^{(T)} \equiv (\pi^{(T)}_t, t \in [T])$, with $\pi^{(T)}_1 \equiv 1$, $\pi^{(T)}_t \equiv \prod_{s \in [t-1]} \bar{p}^{(T)}_s$ for $t \in \{2,\ldots,T\}$; 
\item   $q^{(T)} \equiv (q^{(T)}_t, t \in [T])$, with $q^{(T)}_t \equiv p^{(T)}_t \pi^{(T)}_t$, noting $q^{(T)}_1 = p^{(T)}_1$ and $q^{(T)}_T = \pi^{(T)}_T$; and
\item  $\mu^{(T)} \equiv \sum_{t \in [T]} \pi^{(T)}_t$. 
\end{itemize}
Then: $a)$ $q^{(T)}$ is a probability distribution, $b)$ if $\xsf^{(T)} \sim q^{(T)}$ then $\Pbb(\xsf^{(T)} \geq t) = \pi^{(T)}_t$, and $c)$ $\Ebb[\xsf^{(T)}] = \mu^{(T)}$. 

{\em $a)$}  Set $\Sigma_q^{(T)} \equiv \sum_{t \in [T]} q^{(T)}_t$. It must be shown that $\Sigma_q^{(T)} = 1$.  The proof is by induction in $T$.  The base case $T=1$ is trivial.  Suppose it is true for $T$, i.e., suppose $\Sigma_q^{(T)} = 1$, so that $\Sigma_q^{(T)} = \Sigma_q^{(T-1)} + \pi^{(T)}_T = 1$.  For induction hypothesis case $T+1$:
\begin{eqnarray}
\Sigma_q^{(T+1)}
&=& \sum_{t \in [T-1]} q^{(T+1)}_t + q^{(T+1)}_T + q^{(T+1)}_{T+1} \nonumber \\
&=& \sum_{t \in [T-1]} q^{(T+1)}_t + p^{(T+1)}_T \pi^{(T+1)}_T + \pi^{(T+1)}_{T+1} \nonumber \\
&=& \sum_{t \in [T-1]} q^{(T+1)}_t + p^{(T+1)}_T \pi^{(T+1)}_T + \bar{p}^{(T+1)}_T \pi^{(T+1)}_T \nonumber \\
&=& \sum_{t \in [T-1]} q^{(T+1)}_t + \pi^{(T+1)}_T \nonumber \\
&=& \sum_{t \in [T-1]} q^{(T)}_t + \pi^{(T)}_T = \Sigma_q^{(T)} = 1 
\end{eqnarray}
This proves the induction step.

{\em $b)$}  The proof is by induction in $t$, starting with base case $t=T$, for which the claim holds trivially.  Suppose it holds for $t+1$; it is shown below this implies it holds for $t$:
\begin{eqnarray}
\Pbb(\xsf^{(T)} \geq t) 
&=& \Pbb(\xsf^{(T)} = t) + \Pbb(\xsf^{(T)} \geq t+1) \nonumber \\
&=& q^{(T)}_t + \pi^{(T)}_{t+1} = p^{(T)}_t \pi^{(T)}_t + \pi^{(T)}_{t+1} \nonumber \\
&=& p^{(T)}_t \pi^{(T)}_t + \bar{p}^{(T)}_t \pi^{(T)}_t = \pi^{(T)}_t
\end{eqnarray}
{\em $c)$}  This follows from the elementary fact that the expectation of a nonnegative discrete RV may be expressed in terms of its CCDF as $\Ebb[\xsf] = \sum_t \Pbb(\xsf \geq t)$.  Application to the sequence $\tilde{p}^{\rm SSS}$ with $T = t_{\tilde{p}}^{(1)}$ establishes \eqnref{approxexpunitcostundersss}.
\end{IEEEproof}

\section{Proof of \prpref{epgs}.}
\label{app:epgs}

\begin{IEEEproof}
For finite $(c,x,y)$, 
\begin{equation}
\label{eq:gpye}
\lim_{n\uparrow\infty} n\left[\left(1-\frac{c}{n}\right)^x - \left(1-\frac{c}{n}\right)^{x+y}\right] = yc.
\end{equation}
The three claims from \prpref{epgs} are established in turn.

{\em Claim $i)$: SSR and SSS.} By \facref{ssr} and \thmref{lfeo}:
\begin{eqnarray}
\frac{\tilde{p}^{\rm SSS}_t}{p^{\rm SSR}_t} 
&=& \frac{(n_0^{e,*}-n_0^*)(1-s)^t + n_0^*}{n(1-s)^t - \frac{(1-s)}{s}(1-(1-s)^t)} \bigg/ \frac{n_0^{e,*}}{n}\\
&=& \frac{n (1-s)^t n_0^{e,*}- n (1-s)^t n_0^* + n n_0^*}{n (1-s)^t n_0^{e,*} - \frac{(1-s)}{s} n_0^{e,*} +\frac{(1-s)^{t+1}}{s} n_0^{e,*}}.\label{eq:fpst}
\end{eqnarray}
Recall from \eqnref{gpore} that $n_0^{e,*} = n - (1-s)^{n_0^*}(n-n_0^*)$.  Substitution into \eqnref{fpst} and rearranging gives $\tilde{p}^{\rm SSS}_t/p^{\rm SSR}_t=P(n)/(Q(n) + R(n))$ where, using $s(n) = c/n$, 
\begin{eqnarray}
P(n) &=& n\left(\left(1-\frac{c}{n}\right)^t -\left(1-\frac{c}{n}\right)^{n_0^*+t}\right) \nonumber\\
&+& n_0^*\left(\left(1-\frac{c}{n}\right)^{n_0^*+t} - \left(1-\frac{c}{n}\right)^t\right) +n_0^*\nonumber\\
Q(n) &=& n\left[\left(1-\frac{c}{n}\right)^t -\left(1-\frac{c}{n}\right)^{n_0^*+t} +\frac{1}{c}\left[\left(1-\frac{c}{n}\right)^{t+1}\right.\right. \nonumber\\
&-&\left.\left.\left(1-\frac{c}{n}\right) +\left(1-\frac{c}{n}\right)^{n_0^*+1} -\left(1-\frac{c}{n}\right)^{t+n_0^*+1}\right]\right]\nonumber\\
R(n) &=& n_0^*\left[\left(1-\frac{c}{n}\right)^{t+n_0^*} \right. \nonumber \\
& + & \left. \frac{1}{c}\left(\left(1-\frac{c}{n}\right)^{t+n_0^*+1} -\left(1-\frac{c}{n}\right)^{n_0^*+1}\right)\right].
\end{eqnarray}
Given \eqnref{gpye} it follows that
\begin{equation}
\lim_{n\uparrow\infty} P(n) = (c+1)n_0^*, ~ 
\lim_{n\uparrow\infty} Q(n) = cn_0^*, ~ 
\lim_{n\uparrow\infty} R(n) = n_0^*.
\end{equation}
It follows that $\lim_{n\uparrow\infty} \tilde{p}^{\rm SSS}_t/p^{\rm SSR}_t = \lim_{n\uparrow\infty} P(n)/(Q(n) + R(n)) = 1$. This establishes Claim $i)$.

{\em Claim $ii)$: SSR and SSC.}  Since each SSC sample removes a vertex from $V_0\backslash V_0^{e,*}$ it follows that $p_{t}^{\rm SSC} = n_0^{e,*}/(n-t+1)$, and as such, Claim $ii)$ follows directly from \facref{ssr}.

{\em Claim $iii)$: SSC and SSS.} This follows immediately from Claims $i)$ and $ii)$.
\end{IEEEproof}

\section{Proof of \facref{lcagssr}}
\label{app:lcagssr}

\begin{IEEEproof}
Let $p_t^{\rm SSR} \equiv n^{e,*}_G/n$ denote the probability of success under SSR, with $n^{e,*}_G \equiv |V^{e,*}_G|$.  Recall from \defref{perf} that $\csf_u = \csf^{\rm SSR}_u(G,V^*)$, the random {\em unit} cost, denotes the random number of samples until success, and recall from \facref{ssr} that this quantity has distribution $\mathrm{geo}\left(p_t^{\rm SSR}\right)$.  The random {\em linear} cost under SSR, denoted $\csf^{\rm SSR}_l(G,V^*)$, equals $\csf_l = \xsf_1 + \cdots + \xsf_{\csf_u}$, where $\xsf_t$ is the random {\em extended} degree of the star center in sample $t$.  The expected linear cost is
\begin{eqnarray}
c_l^{\rm SSR}(G,V^*) 
&=& \Ebb[ \csf_l^{\rm SSR}(G,V^*) ] \nonumber \\
&=& \Ebb[ (\xsf_1 + \cdots + \xsf_{\tsf-1}) + \xsf_{\csf_u} ] \nonumber \\
&=& \Ebb[ \Ebb[(\xsf_1 + \cdots + \xsf_{\csf_u-1}) + \xsf_{\csf_u}| \csf_u] ] \nonumber \\
&=& \Ebb[ (\csf_u - 1) \Ebb[\xsf_1|\csf_u]] + \Ebb[\Ebb[\xsf_{\csf_u}|\csf_u] ]. \nonumber
\end{eqnarray}
The first term represents the expected linear cost up until but not including the cost of the final sample, while the second term is the expected cost of the final sample.  The unsuccessful samples are identically distributed, due to replacement.  The expected cost of an unsuccessful sample is $\bar{d}^{e,*}_G$ while the expected cost of a successful sample is $d^{e,*}_G$.  The distribution of the cost of a sample is conditionally independent of the number of samples, conditioned on whether or not the sample is successful or not.  As such, $\Ebb[\xsf_1|\csf_u] =  \bar{d}^{e,*}_G$, $\Ebb[\xsf_{\csf_u}|\csf_u] = d^{e,*}_G$, and $\Ebb[\csf_u] = n/n^{e,*}_G$, yielding \eqnref{clssrag}.
\end{IEEEproof}

\section{Proof of \prpref{lcsscer}.}
\label{app:lcsscer}

\begin{IEEEproof}
Consider an urn with $n$ balls of which $k$ are marked.  Draw $m$ balls uniformly at random and let $\xsf$ be the random number of marked balls drawn, with support $\Smc \equiv \{\max\{0,m-(n-k)\},\ldots,\min\{k,m\}\}$, and distribution $\Pbb(\xsf = l) = \binom{k}{l}\binom{n-k}{m-l}/\binom{n}{m}$ for $l \in \Smc$.  The expectation of $\xsf$ is $\Ebb[\xsf] = \sum_{l \in \Smc} l \Pbb(\xsf = l) = \frac{k m}{n}$. Note this is the same as if the $m$ balls were drawn {\em with} replacement, where $\xsf \sim \mathrm{bin}(m,k/n)$.  This expectation is pertinent in the derivation below of the expected degree when sampling an ER random graph using SSC.  Pick uniformly at random any vertex in $\Gsf_t$, i.e., a vertex that was not selected in the first $t$ draws.  This vertex has a random initial degree in $\Gsf_0$ of $\dsf_0 \sim \mathrm{bin}(n-1,s)$.  Let $\xsf_{t-1}$ be the random number of neighbors that are removed in the first $t-1$ samples, so that its random degree in $\Gsf_t$ is $\dsf_t = \dsf_0 - \xsf_{t-1}$.  Then:
\begin{eqnarray}
\Ebb[\dsf_t] 
&=& \Ebb[\Ebb[\dsf_t|\dsf_0]] \nonumber \\
&=& \Ebb[\Ebb[\dsf_0 - \xsf_{t-1}|\dsf_0]] \nonumber \\
&=& \Ebb[\dsf_0 -\Ebb[\xsf_{t-1}|\dsf_0]] \nonumber \\
&=& \Ebb\left[\dsf_0 -\frac{\dsf_0 (t-1)}{n-1}\right] \nonumber \\
&=& \left(1 - \frac{t-1}{n-1}\right)\Ebb[\dsf_0] \nonumber \\
&=& \left(1 - \frac{t-1}{n-1}\right)(n-1)s \nonumber \\
&=& (n-1)s - (t-1) s = (n-t)s.
\end{eqnarray} 
The proof that $\Ebb[\xsf_{t-1}|\dsf_0] = \frac{\dsf_0 (t-1)}{n-1}$ comes from the discussion above, where the $n-1$ ``balls'' are the potential neighbors in $\Gsf_0$ of the randomly selected vertex, of which the $\dsf_0$ marked ``balls'' are the actual neighbors, and $t-1$ ``balls'' are drawn.    

The event $\{\csf_u \geq t\}$, conditioned on $\nsf^{e,*}_0$, means the first $t-1$ samples fail to find the target.  As graph $\Gsf_{u-1}$ has order $n-u+1$, conditioned on the failure of each previous sample, sample $u \in [t-1]$ fails with probability $1 - \nsf^{e,*}_0/(n-u+1)$.  As these random outcomes are independent:
\begin{eqnarray}
\Pbb(\csf_u \geq t | \nsf^{e,*}_0)
&=& \prod_{u=1}^{t-1} \left(1 - \frac{\nsf^{e,*}_0}{n-u+1} \right).
\end{eqnarray}
The maximum number of samples possible under SSC, $\csf_{\rm max}^{\rm SSC}(\nsf^{e,*}_0)$, is $n - \nsf^{e,*}_0 + 1$. Substitution into \eqnref{eclapprox} yields \eqnref{aelcsscercetso}.
\end{IEEEproof}

\section{Proof of \prpref{lcssser}.}
\label{app:lcssser}

\begin{IEEEproof}
The proof follows the same lines as that of \prpref{lcsscer}, i.e., we establish $i)$ $\Ebb[\dsf_t]$, $ii)$ $\csf_{\rm max}^{\rm SSS} | \nsf^{e,*}_0$, and $iii)$ $\Pbb(\csf_u \geq t | \nsf^{e,*}_0)$.

$i)$ $\Ebb[\dsf_{t}]$.  The expected degree of the star center in sample $t$ is obtained as follows.
Consider any vertex not yet removed by the first $t-1$ samples.  Conditioned on $\nsf_{t-1}$, this vertex has a random degree $\dsf_{t} | \nsf_{t-1} \sim \mathrm{bin}(\nsf_{t-1}-1,s)$, and as such
\begin{equation}
\Ebb[\dsf_t] = \Ebb[\Ebb[\dsf_t|\nsf_{t-1}]] = s\Ebb[\nsf_{t-1}-1],
\end{equation}
By case $ii)$ \eqnref{muWDtc2} in \lemref{watchdraw}:
\begin{eqnarray}
\Ebb[\dsf_t] 
&=& s (\Ebb[\nsf_{t-1}]-1) \nonumber \\
&=& s \left(n (1-s)^{t-1} - \frac{(1-s)}{s}(1-(1-s)^{t-1}) - 1\right) \nonumber \\
&=& n s (1-s)^{t-1} - (1-s)(1-(1-s)^{t-1}) - s
\end{eqnarray}
and as such $\Ebb[\dsf_t^e]$
\begin{eqnarray}
&=& n s (1-s)^{t-1} - (1-s)(1-(1-s)^{t-1}) + (1-s) \nonumber \\
&=& n s (1-s)^{t-1} + (1-s)^t 
= ((n-1)s + 1)(1-s)^{t-1}
\end{eqnarray}

$ii)$ $c_{\rm max}^{\rm SSS}(\nsf^{e,*}_0)$. The maximum number of samples before the target set is reached may be upper bounded by $t^{(1)}_{\tilde{p}}$ the time at which $\tilde{p}_{t+1}^{\rm SSS}=1$ given in \eqnref{t1tildep} of \thmref{lfeo}, i.e. $c_{\rm max}^{\rm SSS}(\nsf^{e,*}_0) \leq t^{(1)}_{\tilde{p}}$. 

$iii)$ $\Pbb(\csf_u \geq t | \nsf^{e,*}_0)$. The probability of requiring $t$ or more samples under SSS is approximated by leveraging the results in \thmref{appuncprobhittarsetSSS}, namely, 
\begin{equation}
\Pbb(\csf_u \geq t | \nsf^{e,*}_0) \approx \prod_{u=1}^{t-1} (1 - \tilde{p}_{u}^{\rm SSS}), 
\end{equation}
where $\tilde{p}_t^{\rm SSS}$ is defined in \eqnref{paucl} in \thmref{lfeo}.

As in the case of \prpref{lcsscer}, \eqnref{aelcsssercetso} follows by substitution.
\end{IEEEproof}

\end{document}